\documentclass[usenatbib]{mn2e}
\usepackage{longtable}
\usepackage{graphicx}
\usepackage{amsmath}
\usepackage{amssymb}
\usepackage{times}
\usepackage{mathtools}
\usepackage{hyperref}
\usepackage{morefloats}
\usepackage{etex}
\pdfoutput=1

\title[G2C2 - III. Structural parameters for Galactic GCs]{
G2C2 - III. Structural parameters for Galactic globular clusters in SDSS passbands}  
\author[J. Vanderbeke et al.]
{Joachim Vanderbeke$^{1,2}$\thanks{E-mail: Joachimvanderbeke@gmail.com}, Roberto De Propris$^3$,  
Sven De Rijcke$^{1}$, Maarten Baes$^{1}$, \newauthor Michael J. West$^{4,2}$, John P. Blakeslee$^{5,6}$
\\ \\
$^{1}$ Sterrenkundig Observatorium, Universiteit Gent, Krijgslaan 281 S9, B-9000 Gent, Belgium\\
$^{2}$ European Southern Observatory, Alonso de C\'{o}rdova 3107, Vitacura, Santiago, Chile\\ 
$^{3}$ Finnish Centre for Astronomy with ESO (FINCA), University of Turku, V{\"a}is{\"a}l{\"a}ntie 20,
   FI-21500 Piikki{\"o}, Finland \\
   $^{4}$ Maria Mitchell Observatory, 4 Vestal Street, Nantucket, MA 02554, USA \\
$^{5}$ Herzberg Insititute of Astrophysics, National Research Council, Victoria, BC V9E2E7, Canada \\
$^{6}$ Department of Physics and Astronomy, Washington State University, 1245 Webster Hall, Pullman, WA 99163-2814, USA \\
}

\defcitealias{Vanderbeke2014a}{Paper~I}  
\defcitealias{Vanderbeke2014b}{Paper~II}

\begin{document}

\date{Accepted. Received }

\maketitle

\label{firstpage}

\begin{abstract}
We use our Galactic Globular Cluster Catalog (G2C2) photometry for 111 Galactic globular clusters (GC) in $g$ and $z$, as well as $r$ and 
$i$ photometry for a subset of 60 GCs and $u$ photometry for 22 GCs, to determine the structural parameters assuming King (1962) models. 

In general, the resulting core radii are in good comparison with the current literature values. However, our half-light radii are slightly lower than the 
literature. The concentrations (and therefore also the tidal radii) are poorly constrained mostly because of the limited radial extent of our imaging. 
Therefore, we extensively discuss the effects of a limited field-of-view on the derived parameters using mosaicked SDSS data, which do not suffer
from this restriction. We also illustrate how red giant branch (RGB) stars in cluster cores can stochastically induce artificial peaks in the surface
brightness profiles. The issues related to these bright stars are scrutinised based on both our photometry and simulated clusters. We also examine 
colour gradients and find that the strongest central colour gradients are caused by central RGB stars and thus not representative for the cluster 
light or colour distribution. 

We recover the known relation between the half-light radius and the Galactocentric distance in the $g$-band, but find a lower slope for redder filters. We did not find a correlation 
between the scatter on this relation and other cluster properties. We find tentative evidence for a correlation between the half-light radii and the
[Fe/H], with metal-poor GCs being larger than metal-rich GCs. However, we conclude that this trend is caused by the position of the clusters in 
the Galaxy, with metal-rich clusters being more centrally located. 

\end{abstract}

\begin{keywords}
Galaxy: Globular Glusters
\end{keywords}

\section{Introduction} \label{sec:intro}

Galactic globular clusters are ancient stellar systems (the majority being formed beyond ${z\sim3}$), containing of $\sim10^5$ stars 
within a volume of $\sim 100$ pc$^3$. They provide essential information on the formation and evolution of the Galaxy \citep[e.g.,][]{Forbes2010} 
and are natural laboratories for theories of stellar structure \citep[and references therein]{VandenBerg2013}. The structure and 
properties of globular clusters bear the imprint of the initial conditions of their formation and interactions with the galactic environment 
\citep{Brodie2006}. The high stellar densities (without a dark matter halo) make globular clusters invaluable objects for the study of N-body 
dynamics \citep[e.g.,][]{Elson1987,Heggie2003,Trenti2010,Hurley2012}. We also expect that these complex environments are involved in 
the formation of several stellar exotica, such as blue stragglers \citep{Ferraro2009,Simunovic2013}, extreme horizontal branch stars 
\citep{FusiPecci1992}, cataclysmic variables and millisecond pulsars \citep{Benacquista2013}, intermediate-mass black
holes \citep{Lutzgendorf2013} and black hole binaries \citep{Lin2013}. 

Structural parameters for globular clusters are needed to explore correlations between stellar populations, dynamics and the Galactic 
environment. The Milky Way is the only object where these questions can be explored in detail, as we can resolve clusters to the level of 
individual stars on the main sequence and study their internal kinematics \citep{Frank2012,Hernandez2013, Bianchini2013,Fabricius2014,
Kacharov2014}. Most structural parameters for Galactic globular clusters are still measured from surface brightness (hereafter SB) profiles 
derived from an inhomogeneous compilation of older CCD and photographic data \citep{Trager1995}. The most extensive compilation 
\citep{McLaughlin2005} fits these to the classical \cite{King1962,King1966} profile for self-gravitating systems  (\citealt{Wilson1975} and
other profiles are also used). Recently, \cite{Miocchi2013} have derived structural parameters based on star counts from Hubble Space 
Telescope (HST) and ground-based photometry for 26 clusters. This latter method is arguably the most accurate \citep[e.g.,][]
{Ferraro1999,Ferraro2003} but it is observationally expensive \citep[e.g.,][]{Salinas2012} and cannot be applied to unresolved 
systems such as clusters in all but the nearest galaxies \citep{Wang2013}. As long as the clusters are bright and well-populated, 
we expect that SB profiles will perform adequately (\citealt{Noyola2006} - but, see also \citealt{Goldsbury2013}). However, in nearly 
all instances (and almost irrespective of methodology) it is difficult to decide on a 'radius' where globular clusters 'end'. In several cases, 
clusters are seen to contain 'extra-tidal' stars (e.g., NGC 1851 -- \citealt{Olszewski2009}; NGC 5694 -- \citealt{Correnti2011}), while tidal tails 
and other debris are relatively common \citep[e.g.,][]{Grillmair1995,Odenkirchen2001,Sollima2011}. A further complication is that about $20\%$ 
of the clusters do not fit smooth King-like models, but exhibit a central density enhancement or core-collapse \citep{Cohn1980}. Given the 
large ages of these objects, this is expected to have occurred in most clusters \citep{Djorgovski1993,Harris1996}, but the formation of hard 
binaries is expected to halt this process \citep{Vesperini1994,Fregeau2007}.

We have recently completed multi-wavelength observations in the Sloan Digital Sky Survey (SDSS) $ugriz$ passbands for the bulk of the 
southern Galactic GCs and supplemented these data with northern clusters from the SDSS DR9 \citep{York2000,Ahn2012}. This results in a
total dataset with $g$ and $z$ photometry for 111 GCs, as well as further $r$ and $i$ photometry for 60 clusters and $u$-band imaging for 22 
GCs. We have presented a study of the integrated $griz$ aperture magnitudes and colours for these objects in \citet[hereafter Paper I]{Vanderbeke2014a} and used these colours 
to improve colour-metallicity relations in \citet[hereafter Paper II]{Vanderbeke2014b}, the first two papers in this series. Here we use the imaging 
to produce SB profiles and derive structural parameters by fitting King models to our homogeneous survey of the Galactic globular cluster system. 
The G2C2 survey has some clear advantages when compared to previous studies: our data are uniform, the measurements are based on CCD 
imaging and the clusters are observed during photometric nights (\citetalias{Vanderbeke2014a}). Moreover, our data cover 
the bulk of the Galactic GCs, were taken in the popular SDSS filter system and were analysed carefully and homogeneously.  

The paper is organised as follows. In Section~\ref{sec:data} we describe the data. Section~\ref{sec:structpars} reports on our analysis, 
presents the structural parameters and compares those with previous work. At this stage we simulate mock clusters to test our fitting 
algorithm and explore possible sources for parameter biases. We discuss the observational biases in more detail and use the large images for the clusters from the SDSS (which theoretically can be studied to any radius) to study the limitations of our dataset 
by artificially constraining the extent of the SDSS SB profiles. We also scrutinise the effects of bright red giant stars on the centering of cluster 
profiles and its consequences for the SB profiles and the King model fits, which further confirm the findings based on the mock data. 
In Section~\ref{sec:complit} we compare our \cite{King1962} parameters with studies from the literature using both SB and count density profiles. In Section~\ref{sec:discussion} we discuss our results and explore their significance for internal colour gradients, stellar populations and the interactions of clusters with our Galaxy. Finally, in Section~\ref{sec:conclusions} we summarise our conclusions. 

\section{Dataset}\label{sec:data}
The dataset for this paper is based on the Galactic Globular Cluster Catalog: a homogeneous imaging survey 
of a large fraction of the Galactic globular cluster system in the SDSS passbands. \citetalias{Vanderbeke2014a} details the data acquisition process and describes 
how these data were reduced. That study also presented our procedures to correct for contamination, how we deal with extinction and calibrate 
photometry. We refer to the above paper for details, but give here a short summary of the essential information. The bulk of the observations 
were carried out on the Cerro Tololo Inter-American Observatory (CTIO) 0.9\,m telescope in a series of observing runs between 2003 and 2013, with most of the data coming from the 
2004 June season. We performed the standard CCD reduction procedures and cosmic ray removal, followed by calibration on the SDSS system
using stars from \cite{Smith2002}. For all clusters we have $g$ and $z$ photometry, often taken on several different nights, while for an important 
subset we also have $r$ and $i$ imaging. We supplemented this with available clusters within the SDSS footprint \citep{Ahn2012}. This resulted 
in a total sample of 111 GCs with $gz$ photometry and 60 GCs with $ri$ photometry. 

In \citetalias{Vanderbeke2014a} we used these data to calculate aperture photometry within the (literature) half-light radius (from the 2010 edition of the 
\citealt{Harris1996} compilation, which is the version we will use throughout this study). This required us to deal with estimating the 
sky level from CTIO images that do not reach to the cluster tidal radius. As this is an important issue for the present paper, we discuss 
this further below. We also adaptively searched for the cluster centre and we similarly describe this in greater detail in this paper. Moreover, 
we experiment in the current study with a new centring method based on RGB stars in Section~\ref{sec:obsbiases}. Finally, we carried
out removal of contaminating (foreground) stars to obtain clean aperture magnitudes and corrected for extinction using the most recent
values from \cite{Schlafly2011}. However, in \citetalias{Vanderbeke2014b} we found some evidence that these extinction values may suffer from systematic 
errors, especially near the plane and the Galactic bulge. 

For several GCs in \citetalias{Vanderbeke2014a} we were unable to carry out aperture photometry because bright red giants in the unresolved cluster centres saturated 
the CCD. Our purpose in the above publication was to provide a conservative set of aperture magnitudes and colours and we therefore excluded 
a number of clusters from our analysis. Here, we fit King models to the surface brightness profiles and we are able to apply an iterative clipping
method and profile fitting to derive 'model' magnitudes. 

\section{Structural parameters}\label{sec:structpars}
\subsection{Surface Brightness Profiles}\label{sec:SBprofiles}
We derive radial surface brightness profiles for all clusters and all passbands by using conventional aperture photometry, in annuli of increasing 
radius, similar to the procedure of building a curve of growth for extended systems. For the CTIO 0.9\,m data, where the field of view is 
$13.6\arcmin$ on the side, the annuli are 15 pixels (corresponding to $\sim5.9\arcsec$) wide, or approximately 4 times the seeing disk. We 
integrated to the edges of the images to obtain the surface brightness profiles to the $\sim6.5\arcmin$ limit allowed by the CTIO 0.9m field of
view. For SDSS data, we can theoretically integrate the profile to infinity, but we also simulate observations limited to the $4.2'$ and $6.5'$ 
field of view used for all other clusters to test the effects of the limited aperture on structural parameters.

We fit the surface brightness profile to a \cite{King1962} model of the form:
\begin{equation} \label{eq:king}
SB(r) = k  \Bigg(\frac{1}{[1+(r/r_c)^2]^{1/2}}-\frac{1}{[1+(r_t/r_c)^2]^{1/2}}\Bigg)^2, 
\end{equation}
where $r$ is the aperture radius, $r_c$ is the core radius and $r_t$ the tidal radius. The constant
$k$ is related to the central surface brightness $\mu$ as 
\begin{equation*} 
\mu = k\Bigg(1-\frac{1}{[1+(r_t/r_c)^2]^{1/2}}\Bigg)^2.
\end{equation*}
We fit these using a Levenberg-Marquardt non-linear least squares algorithm \citep{Press1974,Bevington1992}. We define a concentration 
index $c^\prime=r_t/r_c$ to simplify our numerical work, while $c\equiv \log(r_t/r_c)$ as in \cite{McLaughlin2005}. We iterated the fit with 
5 $\sigma$ clipping about each SB point to remove the effects of bright contaminating stars. We also truncate the SB profiles, excluding the 
SB points which are outside the tidal radius of the previous iteration. Therefore the outer regions of the SB profile were clipped out for some 
very faint clusters, hence the final fit did not rely on the full $6.5\arcmin$ SB profile. The iteration process was stopped when the input SB 
profile vector did not change anymore when compared to the one of the previous loop. As a consequence of the $5 \sigma$ clipping, 
the central SB point was also ignored in some cases, mostly when the SB profile was centred on bright RGB stars. 

\begin{figure*}
\centering
\includegraphics[scale=0.87,trim= 2.8cm 13.2cm 0 8.1cm] {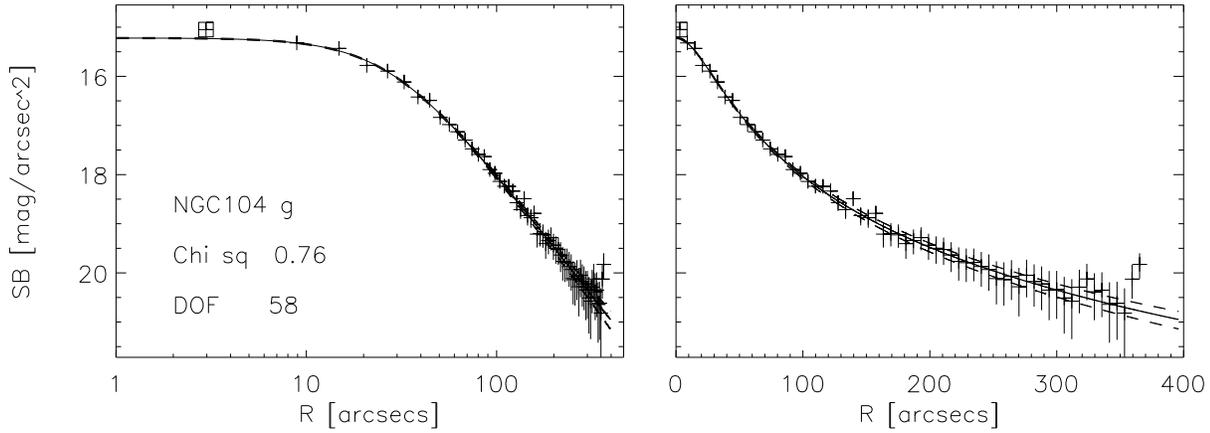}
\caption{SB distribution of NGC~104 in the $g$-band. The left panel focuses on the inner region of the cluster, while the right panel gives details on the outer regions. The dashed lines show the 1-sigma 
deviation of the model, with squares representing the data points that have been clipped iteratively. } 
\label{King_NGC104}
\end{figure*}

We show an example King profile for the classical rich cluster NGC~104 (47 Tucanae) in Fig.~\ref{King_NGC104}. Fits for all other objects 
are placed in an on-line appendix. This rich massive cluster extends much further than our limited FOV, thus the concentration index cannot 
be reliably estimated. It is very difficult in any case to constrain the tidal radii (and the related concentrations) accurately. Since the field of view 
of CTIO data does not reach to the tidal radius for most if not all clusters in the sample, it is inherently difficult to measure this quantity. We 
therefore used an {\it ad hoc} estimate, adopting as the tidal radius the radius at which the best fit King profiles had flux comparable to the mean 
sky noise. For clusters within SDSS, determining the tidal radius is, in principle, not a problem, since one can extrapolate the flux to large
distances; even in this case, the tidal radius is often difficult to determine \citep{Jordan2009}.

\begin{figure*}
\centering 
\includegraphics[scale=0.87,trim= 2.8cm 13.2cm 0 8.1cm] {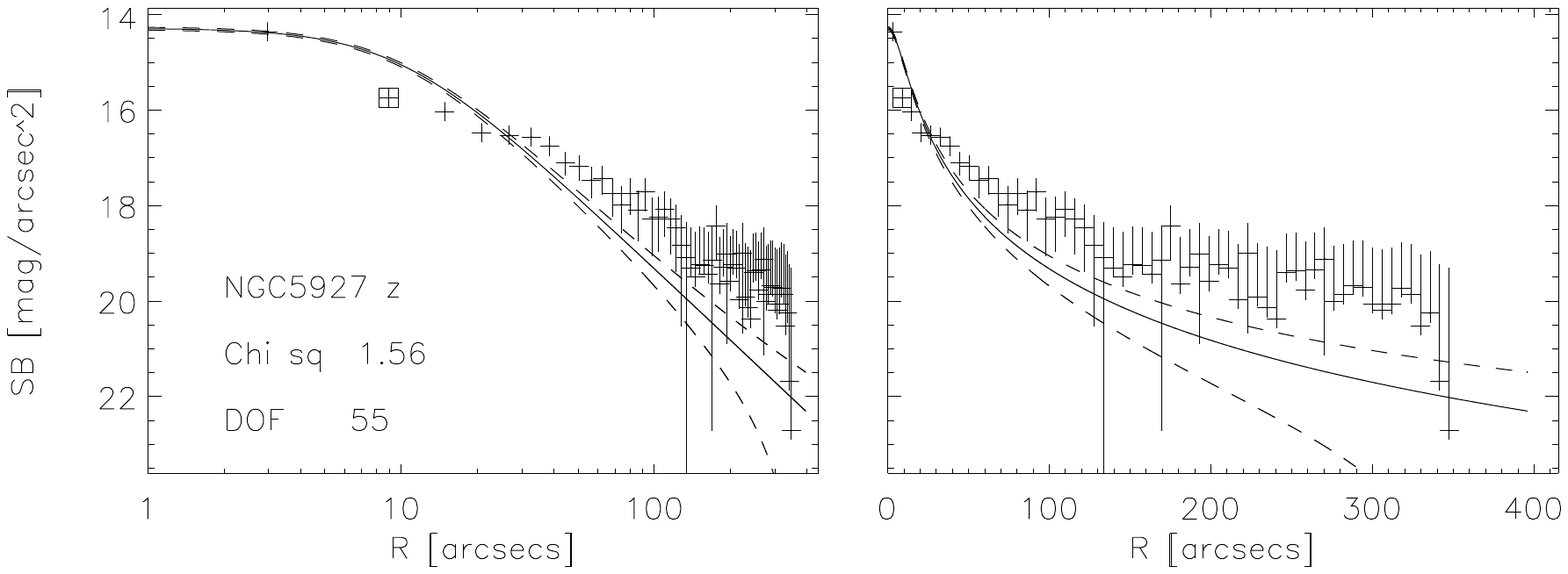} 
\includegraphics[scale=0.87,trim= 2.8cm 13.2cm 0 8.1cm] {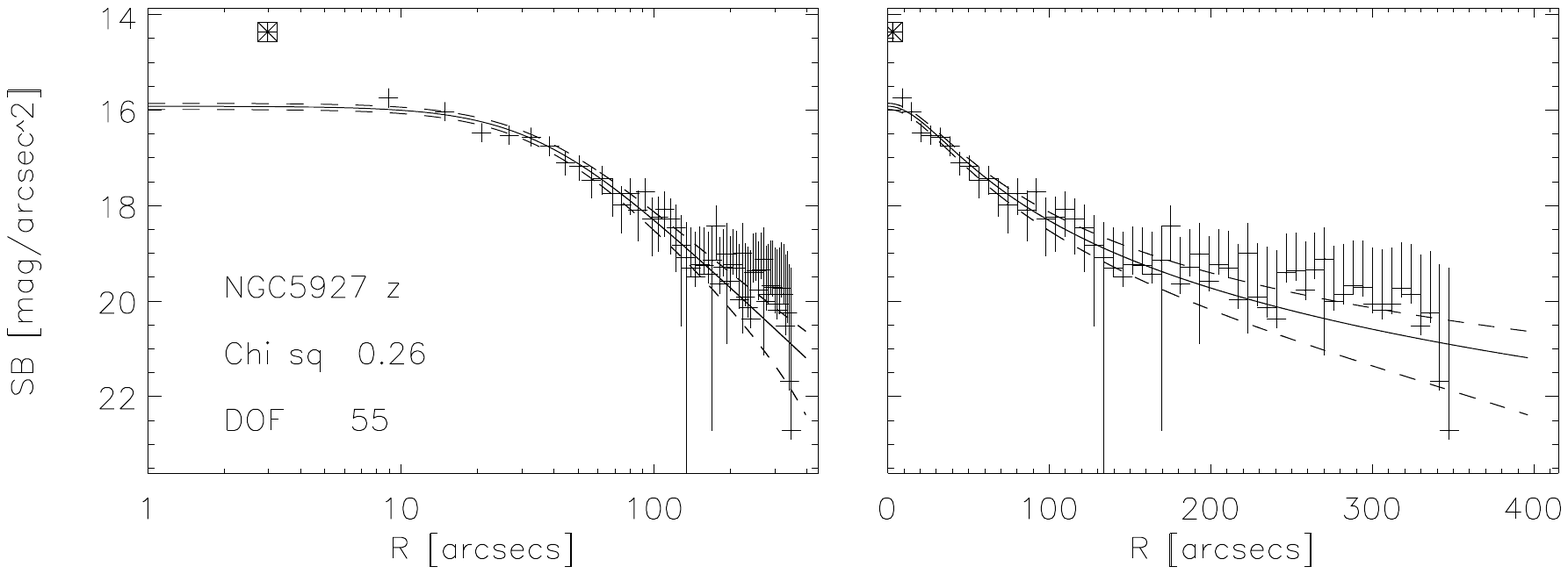}
\caption{Surface brightness distribution for NGC~5927 in the z-band. Legend as in Fig.~\ref{King_NGC104}. The upper panel shows the fit with natural (1/$\sigma$) weighting, the lower panel excludes the central SB point because it is more than 1 magnitude brighter than the 4 adjacent SB points. The asterisk for the central SB point indicates that it was not considered in any of the fitting steps. Boxes indicate the SB points that were not considered in the final fit. }
\label{King_NGC5927_z}
\end{figure*}

We tabulate the derived parameter values based on the CTIO data in Table~\ref{tab:kingpar}. The entire table can be found in the online 
appendix. We estimated the errors on these parameters by bootstrapping the derived fits with 100 random points (assuming Poisson errors) and deriving new King model fits to these artificial data. The errors in the table are the $1\sigma$ conditional errors 
on each parameter. Where we have more than one observation in one or more filters, we always list the King parameters based on the longest 
one in Table~\ref{tab:kingpar} or (for similar observing times) the best reduced $\chi^2$. Note that in some cases (e.g., core-collapsed systems) 
we were forced to exclude the inner annulus from our analysis as it would otherwise drive the whole fit because of its small errors: we exclude the central SB point when it was more than 1 magnitude brighter than the four adjacent SB points. 
One example is  NGC~5927 in the upper panel of Fig.~\ref{King_NGC5927_z}. This could 
bias the results for core-collapsed clusters. However, these clusters are by definition not well represented by King models in any case.

\begin{table*}
\centering
\caption{\label{tab:kingpar} Extract of the King parameter table based on CTIO SB profiles. Central SB uncertainties are pure bootstrapping uncertainties and do not include calibration uncertainties, neither the systematic error introduced in . We do not list the King model concentration $c$, because this parameter could not be estimated reliably based on our data. 
Table~\ref{tab:systerr} presents the systematic errors on each of the parameters. Model and aperture (Aper) magnitudes are computed within the listed $r_h$. In case of multiple observations, the longest observation was chosen (with exposure time ExpT). The reduced $\chi^2$ and the degrees of freedom (DOF) of the fit are also given. $SB0_{in}$ indicates if the central SB point was included by the final fit.}
\begin{tabular}{lccccccccccccc}
\hline
ID & & $\mu_0$ & $\sigma(\mu_0)$& $r_c$ & $\sigma(r_c)$  &$r_h$ & $\sigma (r_h)$ &Model & Aper & ExpT & $\chi^2$ & DOF & $SB0_{in}$ \\
&  & $[\text{mag}/\arcsec^2]$ & & $[\arcmin]$ & & $[\arcmin]$ &&[mag] &[mag]  &[s]&&\\ \hline 
  
         NGC104   &     g   &           14.94   &            0.00   &            0.47   &            0.00   &            2.62   &            0.01   &            5.09   &            5.08   &    60   &            0.76   &    58   &     0   \\
         NGC104   &     z   &           13.82   &            0.00   &            0.51   &            0.00   &            2.52   &            0.01   &            3.89   &            3.85   &    60   &            1.32   &    56   &     1   \\
         NGC288   &     g   &           20.31   &            0.00   &            1.76   &            0.03   &            2.19   &            0.02   &            9.14   &            9.12   &   270   &            0.99   &    58   &     0   \\
         NGC288   &     r   &           19.91   &            0.01   &            2.03   &            0.07   &            2.05   &            0.03   &            8.76   &            8.70   &    60   &            0.57   &    58   &     0   \\
         NGC288   &     i   &           19.68   &            0.01   &            2.32   &            0.12   &            2.01   &            0.01   &            8.49   &            8.41   &    60   &            0.55   &    58   &     0   \\
         NGC288   &     z   &           19.56   &            0.01   &            2.51   &            0.17   &            1.88   &            0.02   &            8.44   &            8.39   &    60   &            0.21   &    58   &     0   \\
         NGC362   &     g   &           15.07   &            0.00   &            0.18   &            0.00   &            0.97   &            0.01   &            7.39   &            7.35   &    60   &            0.44   &    60   &     1   \\
         NGC362   &     r   &           14.39   &            0.00   &            0.17   &            0.00   &            0.80   &            0.01   &            6.89   &            6.92   &    60   &            0.48   &    59   &     1   \\
         NGC362   &     i   &           14.08   &            0.00   &            0.17   &            0.00   &            0.83   &            0.01   &            6.54   &            6.58   &    60   &            0.56   &    59   &     1   \\
         NGC362   &     z   &           13.90   &            0.00   &            0.18   &            0.00   &            0.70   &            0.01   &            6.45   &            6.50   &    60   &            0.28   &    59   &     1   \\
\hline 
\end{tabular} \\
\end{table*}

We now integrate the King profiles numerically to compute a 'model' magnitude within the half-light radius. These are also given in
Table~\ref{tab:kingpar}. For SDSS clusters we give model and aperture magnitudes in Table~\ref{tab:kingparSDSS}: the full
table is available online, while an extract is shown here for guidance. In \citetalias{Vanderbeke2014a}, we presented aperture magnitudes based on the 
half-light from \cite{Harris1996}. Here we determine the half-light radius and we therefore choose to compare the model and aperture 
magnitudes. Generally, these magnitudes agree well, with a median magnitude difference (model$-$aperture) of $\sim0.01\text{\,mag}$.

\begin{table*}
\centering
\caption{\label{tab:kingparSDSS} Extract of the GC King parameters and errors for SB profiles based on SDSS data. }

\begin{tabular}{lcccccccccccc}
\hline
ID & & $\mu_0$ & $\sigma(\mu_0)$& $r_c$ & $\sigma(r_c)$  &$r_h$ & $\sigma (r_h)$ &Model & Aper &  $\chi^2$ & DOF & $SB0_{in}$ \\
&  & $[\text{mag}/\arcsec^2]$ & & $[\arcmin]$& & $[\arcmin]$ &&[mag] &[mag]  &&&\\ \hline 

             NGC2419   &     u   &           20.95   &            0.01   &            0.34   &            0.05   &            0.60   &            0.04   &           12.78   &           12.76  & $           0.05$ & $   49 $ &     1   \\
             NGC2419   &     g   &           19.79   &            0.00   &            0.36   &            0.00   &            1.15   &            0.00   &           10.91   &           10.99  & $           0.76$ & $   52 $ &     1   \\
             NGC2419   &     r   &           19.33   &            0.00   &            0.36   &            0.00   &            1.22   &            0.00   &           10.41   &           10.48  & $           0.61$ & $   52 $ &     1   \\
             NGC2419   &     i   &           19.10   &            0.00   &            0.37   &            0.00   &            1.19   &            0.00   &           10.17   &           10.23  & $           0.78$ & $   52 $ &     1   \\
             NGC2419   &     z   &           19.01   &            0.01   &            0.37   &            0.01   &            0.80   &            0.01   &           10.45   &           10.47  & $           0.08$ & $   50 $ &     1   \\
             NGC5024   &     u   &           18.89   &            0.00   &            0.45   &            0.00   &            0.85   &            0.01   &           10.11   &           10.11  & $           0.06$ & $   42 $ &     1   \\
             NGC5024   &     g   &           17.59   &            0.00   &            0.41   &            0.00   &            1.07   &            0.01   &            8.68   &            8.68  & $           0.73$ & $   52 $ &     1   \\
             NGC5024   &     r   &           17.10   &            0.00   &            0.38   &            0.00   &            1.10   &            0.01   &            8.26   &            8.26  & $           1.23$ & $   53 $ &     1   \\
             NGC5024   &     i   &           17.19   &            0.00   &            0.45   &            0.00   &            1.08   &            0.01   &            8.16   &            8.15  & $           0.67$ & $   53 $ &     1   \\
             NGC5024   &     z   &           16.67   &            0.00   &            0.39   &            0.00   &            0.92   &            0.01   &            7.97   &            7.96  & $           0.59$ & $   50 $ &     1   \\
             NGC5053   &     u   &           23.52   &            0.04   &            2.32   &            0.96   &            1.06   &            0.01   &           13.41   &           13.37  & $           0.01$ & $   56 $ &     1   \\
             NGC5053   &     g   &           22.43   &            0.01   &            2.65   &            0.31   &            2.34   &            0.03   &           10.91   &           10.88  & $           0.25$ & $   56 $ &     1   \\
             NGC5053   &     r   &           22.09   &            0.01   &            2.59   &            0.31   &            2.24   &            0.02   &           10.65   &           10.65  & $           0.30$ & $   51 $ &     1   \\
             NGC5053   &     i   &           21.88   &            0.01   &            2.79   &            0.29   &            2.41   &            0.05   &           10.27   &           10.24  & $           0.36$ & $   56 $ &     1   \\
             NGC5053   &     z   &           21.81   &            0.02   &            2.93   &            0.87   &            1.65   &            0.01   &           10.80   &           10.74  & $           0.10$ & $   56 $ &     1   \\

\hline 
\end{tabular} \\
\end{table*}

\subsection{Sky values}

In order to follow globular cluster profiles to low surface brightness levels, an accurate estimate of the sky value is necessary. Small errors in 
sky values propagate because of the large areas of the apertures, especially in the outer regions of each cluster. We have used the
\textsc{mmm} (Mean, Median, Mode) algorithm to measure the sky flux in apparently blank regions in the corners of the CCD images: the 
routine is adapted from \textsc{daophot} \citep{Stetson1987}, specifically developed for crowded fields as in our globular clusters. The 
algorithm clips sky pixels well above the median and combines median, mean and mode sky values to obtain a more accurate background 
estimate. In our case, exposure times are relatively short to preserve the dynamic range between the bright central regions and the low 
surface brightness wings of the cluster profile; this leads to more uncertain sky determinations (see \citetalias{Vanderbeke2014a} for more details). One can 
compare this with the most commonly used approach to derive SB profiles for galaxies \citep{Peng2002,Peng2010}: in some cases the 
outskirts of extended systems are barely above the sky noise and even the areal increase in the apertures is offset by the increasing 
noise from the sky, flat field and detector read. This may lead to a bias in favour of low effective radii. The problem may be further 
complicated by our assumption of circular symmetry, whereas there is evidence of changing ellipticity and position angles in the outer 
regions of some globular clusters \citep{Bianchini2013}. Additionally, some clusters have stars beyond the tidal radius, while others under-fill 
their tidal region \citep{Gieles2010,Alexander2013}. \cite{Peng2002,Peng2010} argues for the solution we have eventually chosen, to estimate 
the sky independently and hold it fixed for the estimate. Finally, in some cases, some very bright (sometimes saturated) foreground stars 
were replaced by the sky value. Generally, these stars were already clipped by our fitting algorithm. However, the model fits and 
residuals look nicer when masking these saturating foreground stars. It is particularly useful to do this for faint sparse clusters, which are 
sometimes not much brighter than the sky itself. 

\subsection{Centering errors}\label{sec:obsbiases}

An essential element in our procedure is the choice of an appropriate centroid for the apertures. In \citetalias{Vanderbeke2014a} we determined the cluster
centre by using a series of small apertures surrounding the optical centre and choosing the position where the flux is maximal \citep{Noyola2006,
Bellazzini2007}. As a first approximation, this is our estimate of the position of the cluster centre.  During the analysis of the data, it became clear 
that a few bright stars (mostly RGB stars) can strongly affect the derived parameters, mostly because of their effect on the position of the cluster 
centre \citep{Goldsbury2013} and especially for poorer systems. Fitting the centres of GCs accurately has basically been an issue ever since 
King models were fitted to SB profiles. \cite{Djorgovski1986} found that about one-fifth of GC cores are not well fit even with high-concentration 
King models. Collapsed cores, which are better represented by power-law profiles, are encountered frequently. We tested this by creating some 
mock clusters using \textsc{iraf}\footnote{IRAF is distributed by the National Optical Astronomy Observatory, which is operated by the Association 
of Universities for Research in Astronomy under cooperative agreement with the National Science Foundation.}.

\begin{figure}
\centering 
\includegraphics[scale=0.87,trim= 2.7cm 13.2cm 0 5.8cm]{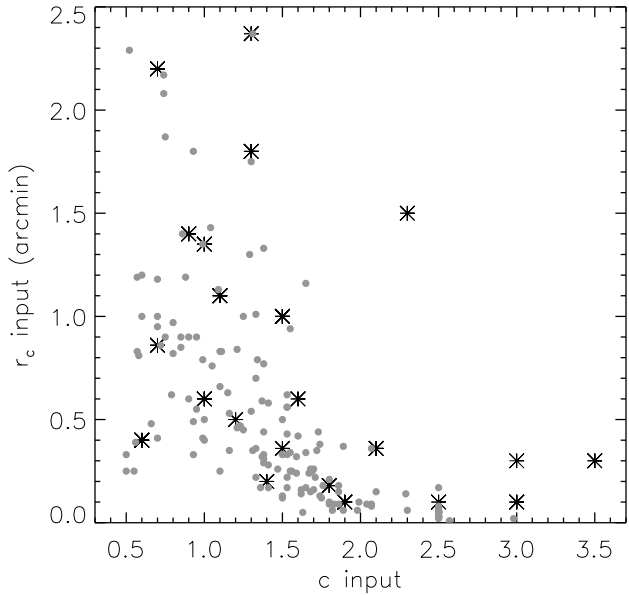}
\caption{Distribution of the input $r_c$ and $c$ parameters for the \textsc{iraf} mock data. Black asterisks represent our input parameters, small filled grey circles show the literature values \citep{Harris1996}.}\label{fig:distr_rc_c}
\end{figure}

Stars were generated using the \cite{Bahcall1980} stellar luminosity function and distributed over the CTIO FOV according to the spatial 
probability distributions determined by different King model density distributions. Fig.~\ref{fig:distr_rc_c} shows the distribution of the input $r_c$ and $c$ in the \textsc{iraf} procedure. These input parameters cover the known values from the literature \citep{Harris1996} well and are extended to somewhat higher concentrations. Because the sampling of the clusters can affect the SB 
profile determination, two approaches were used for the stellar densities: images of well-populated clusters are two-dimensional projections 
of clusters with $10^5$ stars, while more sparse clusters only contain $10^4$ stars. Obviously, the resulting observed density of the cluster 
does depend on $r_c$ and $c$ itself. No foreground or background contamination is included. Exposure times, which determine the 
signal-to-noise of the mock data, were chosen to match our real observations. To derive the cluster centre, we used the above described
procedure from  \cite{Noyola2006} and \cite{Bellazzini2007}.

\begin{figure}
\centering 
\includegraphics[scale=0.87,trim=2.7cm 13.2cm 0 5.8cm]
{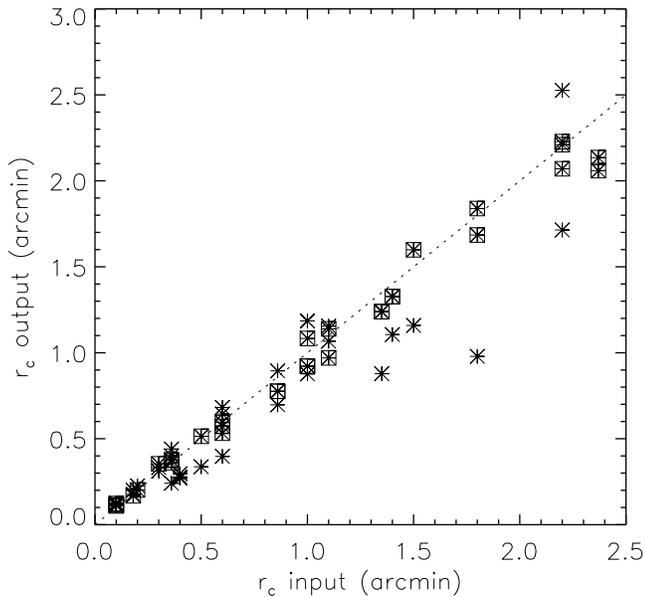}
\caption{Comparison of the input core radii $r_c$ with the resulting $r_c$ based on our fitting algorithm. Well-populated clusters ($N=10^5$) are indicated with a box, other clusters are more sparse ($N\leq10^4$). The dotted line shows the one-to-one correspondence. }\label{fig:mockrc}
\end{figure}

For well-populated clusters measured core radii agree well with the input values (Fig.~\ref{fig:mockrc}). For sparser clusters, the agreement
is worse, with some showing large offsets. A strong outlier (input $r_c\sim1.8\arcmin$, output $r_c\sim1\arcmin$) shows how stochastic effects
may affect our results. We generated this mock GC with $10^4$ stars, hence it is sparsely populated. In fact, the SB profile was centred on a bright 
star, resulting in a strong artificial central peak. This led to a large overestimate of the concentration index. If we exclude, arbitrarily, the central 
SB point, this yields more reasonable values. 

\begin{figure}
\centering 
\includegraphics[scale=0.87,trim=2.7cm 13.2cm 0 5.8cm]
{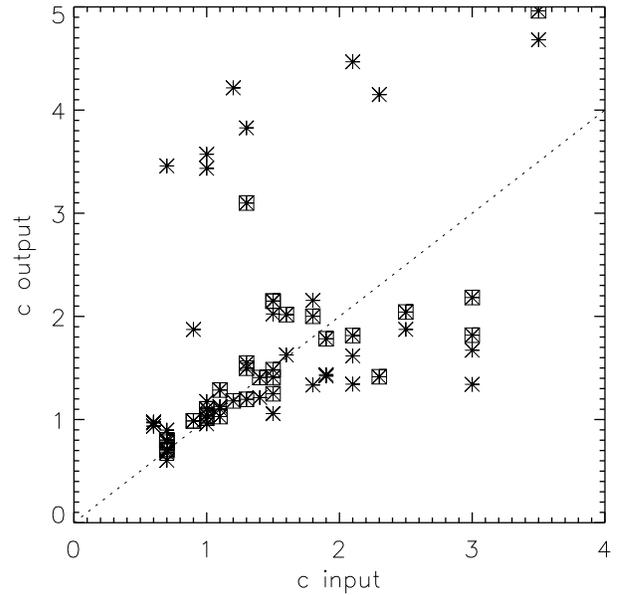}
\caption{Comparison of the \textsc{iraf} input concentrations with the resulting output $c$ based on our fitting algorithm. Legend as in Fig.~\ref{fig:mockrc}. Due to the limited FOV, the concentrations can be overestimated: $c\gtrsim2.5$ are not reliable. See text for more details.} 
\label{fig:mockc}
\end{figure}

In Fig.~\ref{fig:mockc} we compare the \textsc{iraf} input concentrations with the output of the fitting algorithm. We find that a significant fraction of the simulated clusters show large concentration discrepancies, especially for sparsely populated clusters. 
For GCs with tidal radii much beyond the FOV, it is impossible to constrain the concentrations, which is not surprising. In general, the output concentrations larger than about 2.5 are not reliable.

Because the distribution of bright stars is stochastic and the central surface brightness distributions of King models are flat, errors in the centering 
may lead to artificially flat profiles \citep{Mackey2003}. Sometimes, however, uncertainties in determining the cluster center can also produce 
artificially {\it peaked} profiles. Therefore, we also experimented with an alternative centring method, using the weighted mean position of cluster 
red giants to determine the centre of the cluster. We expect that the red giants stars are equally distributed around the centre, hence this method
should not be biased by the stochastic positions of some bright stars. Because this method removes 'artificial' peaks or dips in the SB profile 
\citep{Miocchi2013}, it is found that the core radii increase with respect to our original method.

The following figures compare the core, half-light and central surface brightness derived using our original approach and a centre based
on the positions of red giant stars. The half-light radii compare well between the two approaches, while the central surface brightness decreases
as artificial peaks in the profile are removed.

\begin{figure}
\centering 
\includegraphics[scale=0.87,trim= 2.7cm 13.2cm 0cm 5.8cm]{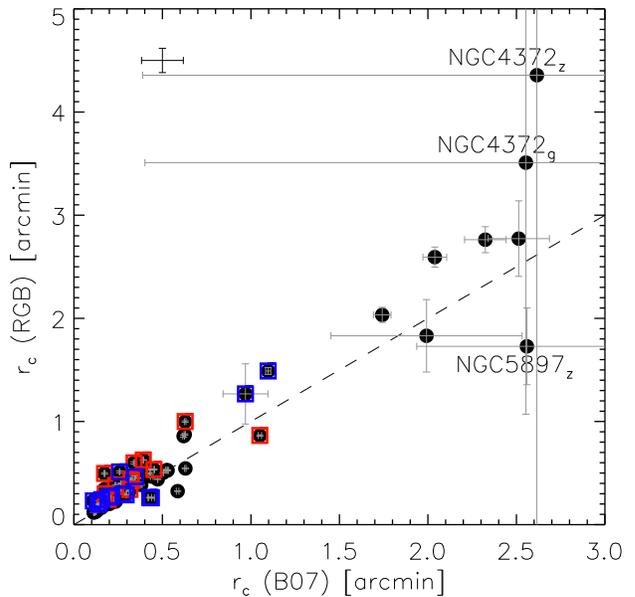}
\caption{Comparison of the core radii $r_c$ based on SB profiles obtained with the roaming procedure \citep[][B07]{Bellazzini2007} and RGB-based centres. The red squares indicate the 'red core' clusters (with $\Delta_{g-z}>0.1$, defined in Section~\ref{sec:colgrad}). Blue squares indicate the 'blue core' clusters (with $\Delta_{g-z}<-0.2$).  The black error bars in the top-left corner show the $z$-band systematic error. See text for more details.  }
\label{fig:rc_roam_RGB}
\end{figure}

\begin{figure}
\centering 
\includegraphics[scale=0.87,trim= 2.7cm 13.2cm 0cm 5.8cm]{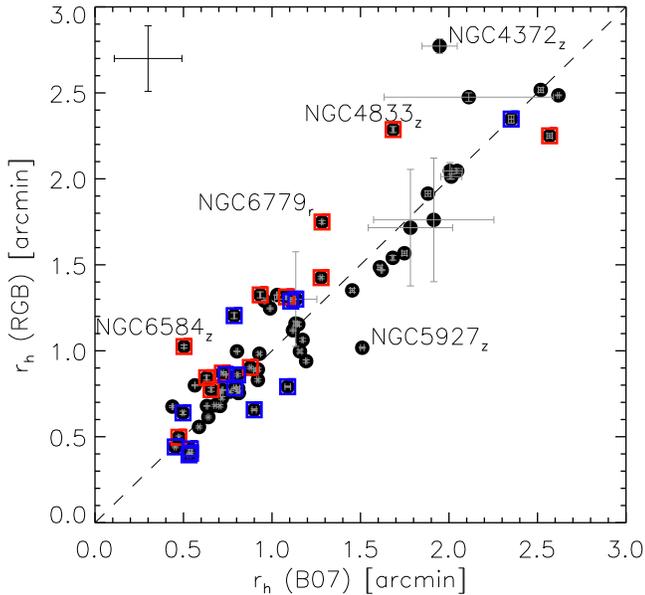}
\caption{Comparison of the half-light radii $r_h$ based on SB profiles obtained with the roaming procedure \citep[][B07]{Bellazzini2007} and RGB-based centres. Legend as in Fig.~\ref{fig:rc_roam_RGB}. See text for more details.  }
\label{fig:rh_roam_RGB}
\end{figure}

\begin{figure}
\centering 
\includegraphics[scale=0.87,trim= 2.7cm 13.2cm 0cm 5.8cm]{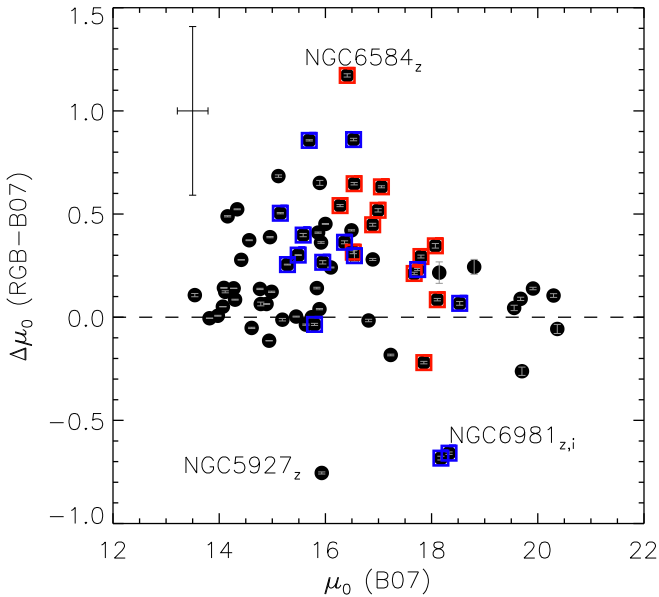}
\caption{Comparison of the central SB based on SB profiles obtained with the roaming procedure \citep[][B07]{Bellazzini2007} and RGB-based centres. Legend as in Fig.~\ref{fig:rc_roam_RGB}. See text for more details. }
\label{fig:mu0_roam_RGB}
\end{figure}

From this, we estimate the systematic error on the parameters introduced by the uncertainty in determining the cluster centre and list the corresponding values in Table~\ref{tab:systerr}. 
\begin{table}
\centering
\caption{\label{tab:systerr} Systematic errors for the structural parameters due to the stochastically distributed RGB stars and the uncertainty on the centre determination. For the $ri$ parameters, the systematic uncertainties for the $z$-band should be adopted for all parameters. See text for more details.}
\begin{tabular}{lccc}
\hline
Filter &  $\sigma_{syst}(\mu_0)$& $\sigma_{syst}(r_c)$ &  $\sigma_{syst}(r_h)$ \\
& $[\text{mag}/\arcsec^2]$ & $[\arcmin]$  & $[\arcmin]$ \\
\hline 
g & 0.228 &  0.077  &   0.101	 \\
z & 0.289  & 0.118     &  0.191		\\
\hline
\end{tabular} \\
\end{table}

\subsection{Effects of a limited field of view}\label{sec:GCCTIOSDSS}

We compare clusters in common between the SDSS and our CTIO data. In principle, there is no limit to how far a cluster can be
followed with SDSS photometry and therefore we can test how our limited FOV affects model parameters. Note, however, that even 
SDSS data have their limitations: bright RGB stars can saturate the SDSS CCD; moreover, the mosaicking of the SDSS data can result 
in artificial sky gradients.

\begin{figure*}
\centering 
\includegraphics[scale=0.9,trim= 2.7cm 13.2cm 0 5.5cm]
{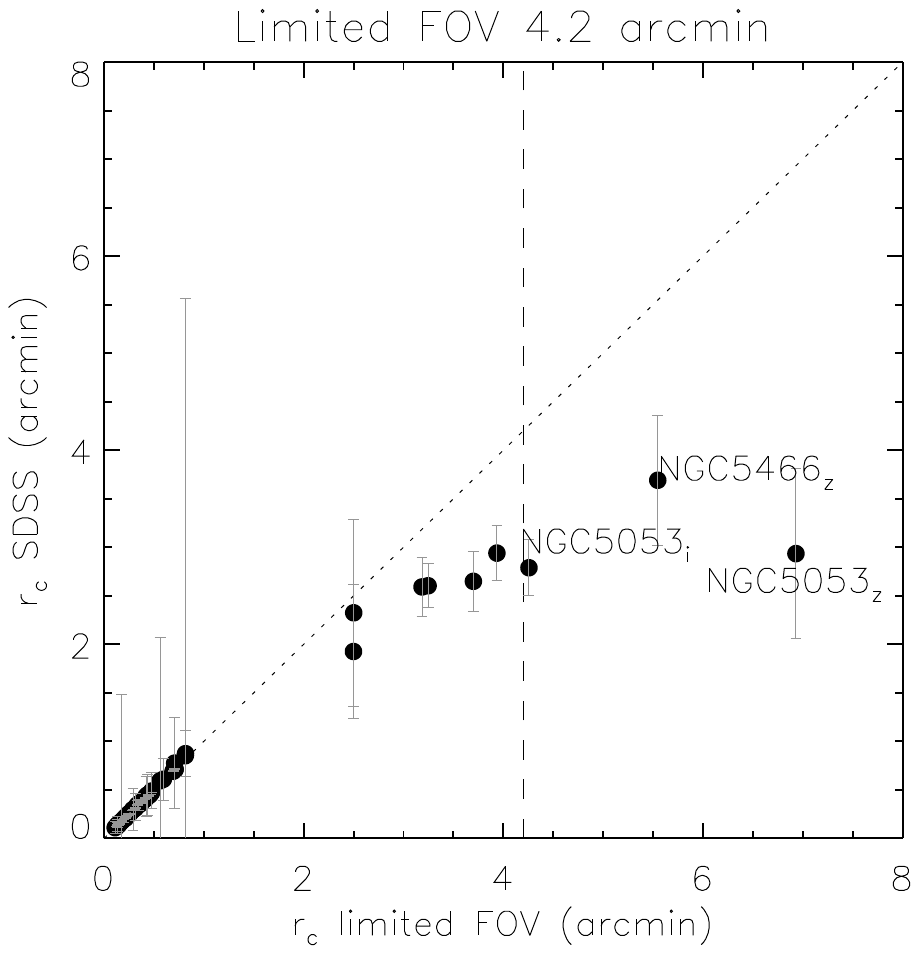}\includegraphics[scale=0.9,trim= 11.5cm 13.2cm 0 5.5cm]
{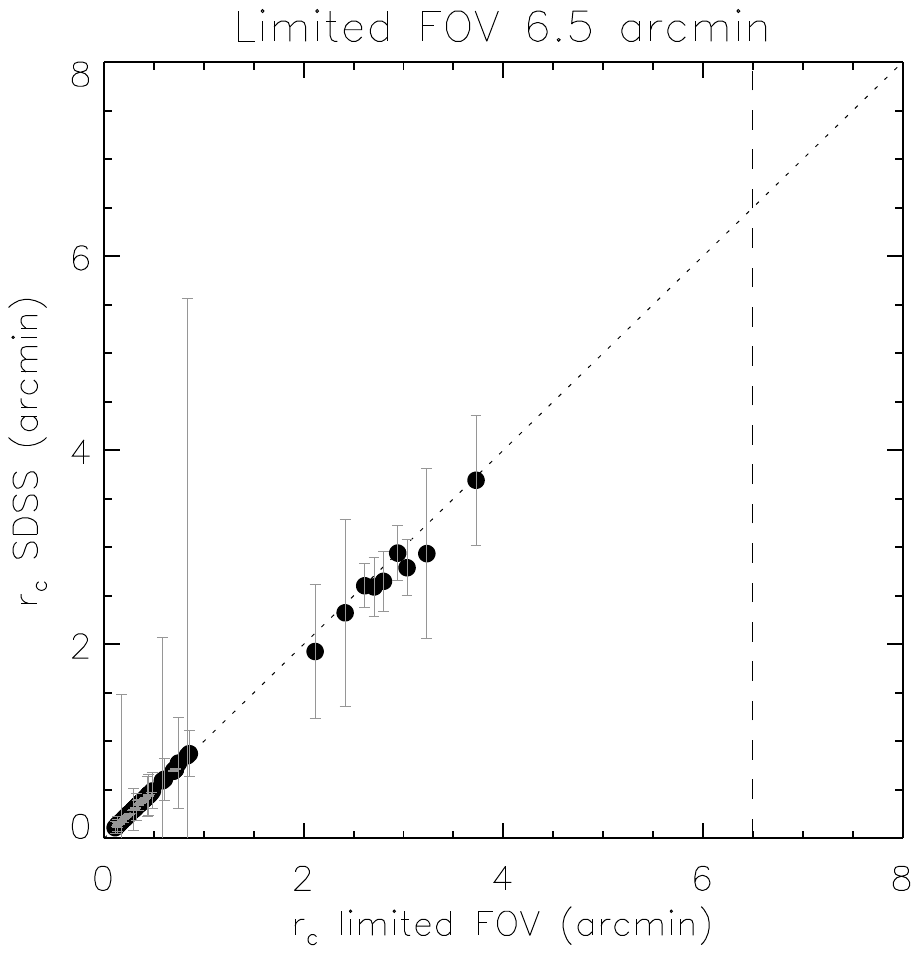}
\caption{Comparison of the core radii $r_c$ based on the full mosaicked FOV and artificially limited SB profiles of $4.2\arcmin$ and $6.5\arcmin$ (left and right panel, respectively) for all SDSS GCs belonging to the clean sample introduced in Section~\ref{sec:complit} (e.g. 19 GCs for $g$, 18 GCs for $z$). The artificial cuts limiting the SB profile radius are indicated with the dashed lines. The dotted line represents the one-to-one correspondence. It is clear that the scatter is significantly reduced using the 6.5$\arcmin$ SB profiles, stressing the importance of using the entire CTIO FOV. See text for more details. }
\label{fig:fake_limited_FOV}
\end{figure*}

We use the SDSS mosaics to determine the surface brightness profile and fit this to a King model. We then artificially restrict the field to 
radii of $4.2\arcmin$ 
and $6.5\arcmin$ (the total CTIO FOV), to estimate the influence of the field of 
view on the model fits. In general, we find that the smaller field of view biases the results towards a smaller core radius, but using the 
full $6.5\arcmin$ FOV solves this issue (Fig.~\ref{fig:fake_limited_FOV}). Note that tidal tails were found in NGC~5053 \citep{Jordi2010}.

\begin{figure*}
\centering 
\includegraphics[scale=0.9,trim= 2.7cm 12.9cm 0 5.5cm]
{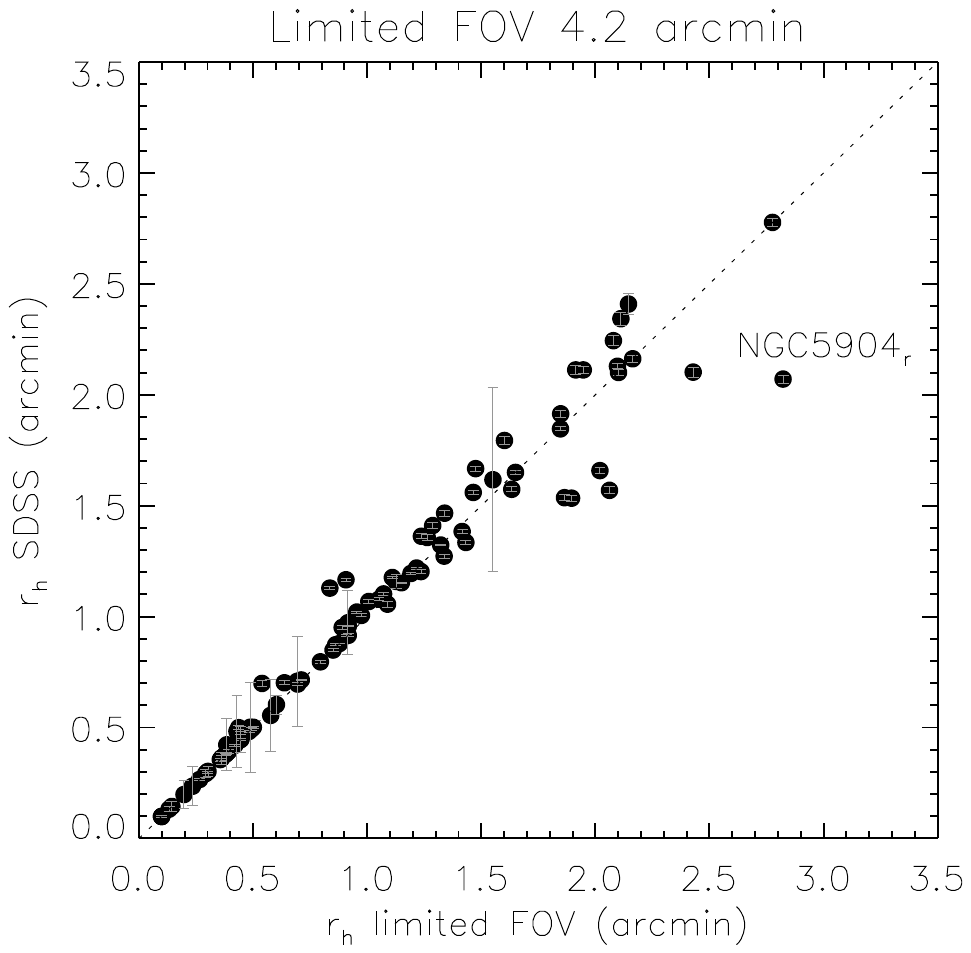}\includegraphics[scale=0.9,trim= 11.5cm 12.9cm 0 5.5cm]
{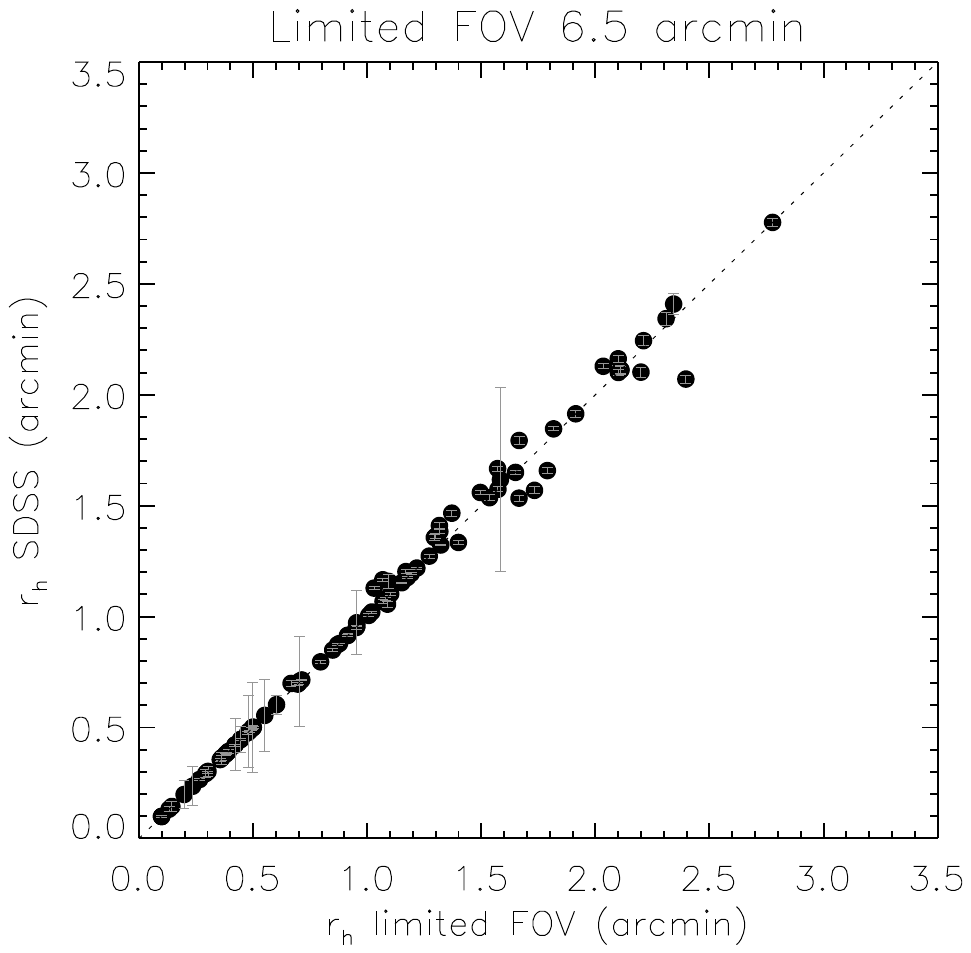}
\caption{Comparison of the half light radii $r_h$ based on the full mosaicked FOV and an artificial limited SB profile for the SDSS GCs belonging to the clean sample introduced in Section~\ref{sec:complit} (e.g. 19 GCs for $g$, 18 GCs for $z$). The dotted line represents the one-to-one correspondence. It is clear that the scatter is reduced using the 6.5$\arcmin$ SB profiles, stressing the importance of using the entire CTIO FOV. See text for more details. }
\label{fig:rh_fake_limited_FOV}
\end{figure*}

In Fig.~\ref{fig:rh_fake_limited_FOV} we present a comparison of the half light radii based on the 
full mosaicked SDSS frame and the artificially truncated profiles. These two quantities compare well, which suggests that the half-light radius is also generally well determined. The scatter on the one-to-one correspondence is again reduced when using the $6.5\arcmin$ SB profile instead of the $4.2\arcmin$ SB profile. 

\begin{figure*}
\centering 
\includegraphics[scale=0.6,trim= 2.3cm 13.2cm 9.5cm 6cm]{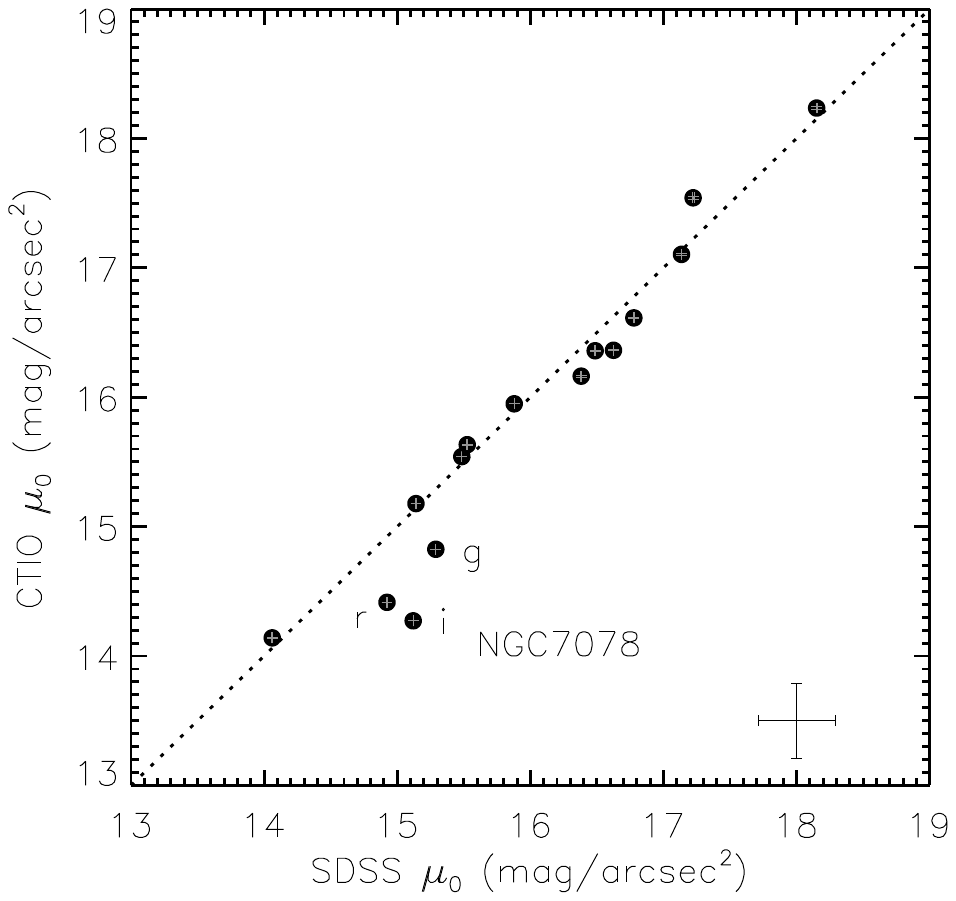}\includegraphics[scale=0.6,trim= 2.3cm 13.2cm 9.5cm 6cm]{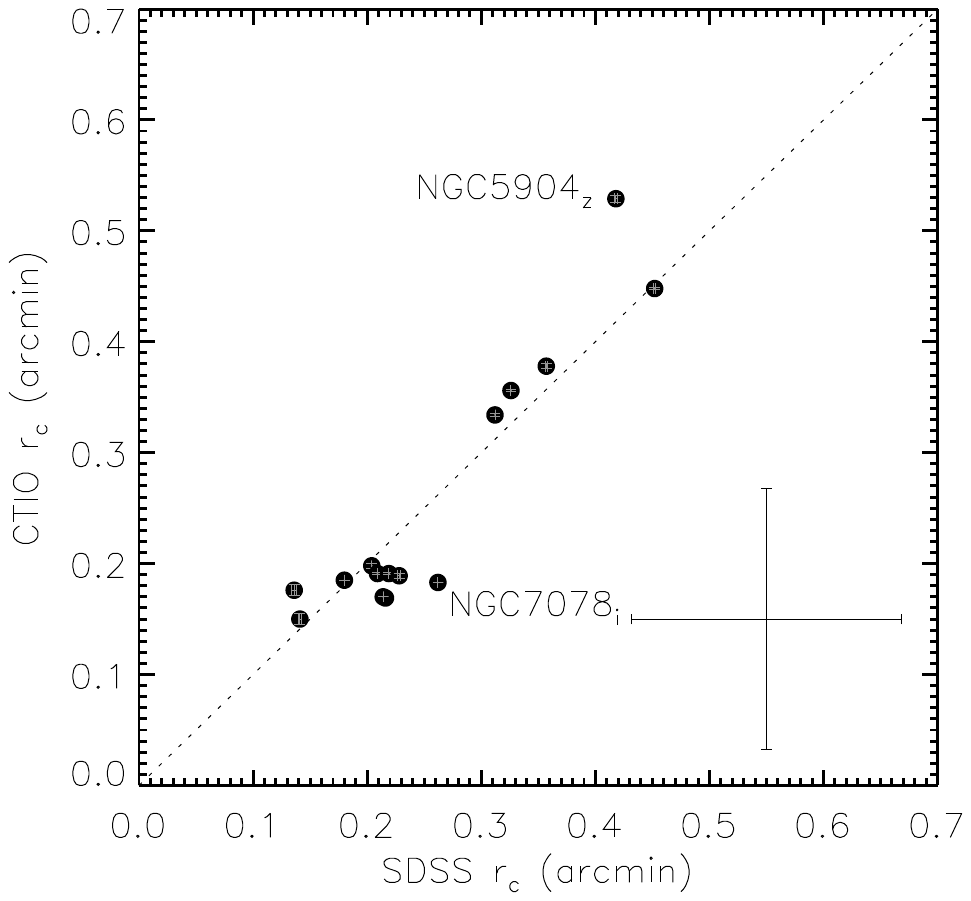}\includegraphics[scale=0.6,trim= 2.3cm 13.2cm 9.5cm 6cm]{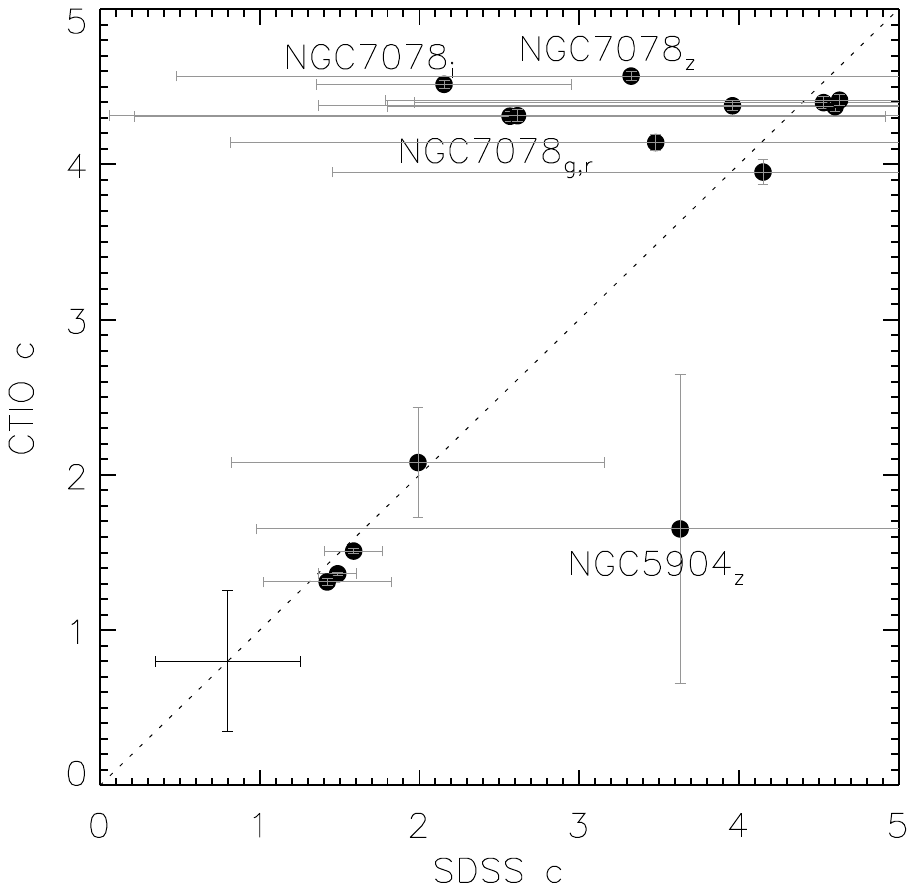}
\caption{Comparison of the King model parameters (central SBs $\mu_0$, core radii $r_c$ and concentrations $c$) based on the full mosaicked SDSS FOV and on the CTIO data. The dotted line represents a one-to-one correspondence. The $z$-band systematic errors are shown by black error bars. } \label{fig:SDSS_CTIO}
\end{figure*}

Fig.~\ref{fig:SDSS_CTIO} shows a comparison of the structural parameters for clusters with both SDSS (full images) and CTIO data. 
The central surface brightnesses are in good agreement (with one exception -- NGC~7078 -- whose profile is notoriously difficult to fit -- \citealt{
Newell1978}). Core radii are also in good agreement, while concentrations from our data are generally found to be unreliable.

\subsection{Comparison of observations in the $gz$ filters}\label{sec:appendix_compare_gz}

Different filters are sensitive to different stars. However, if the King parameters truly reflect the star count densities, we do not expect 
substantial differences between the structural parameters based on observations in the blue or red. 

\begin{figure*}
\centering 
\includegraphics[scale=0.6,trim= 2.7cm 13.2cm 9cm 5.8cm]{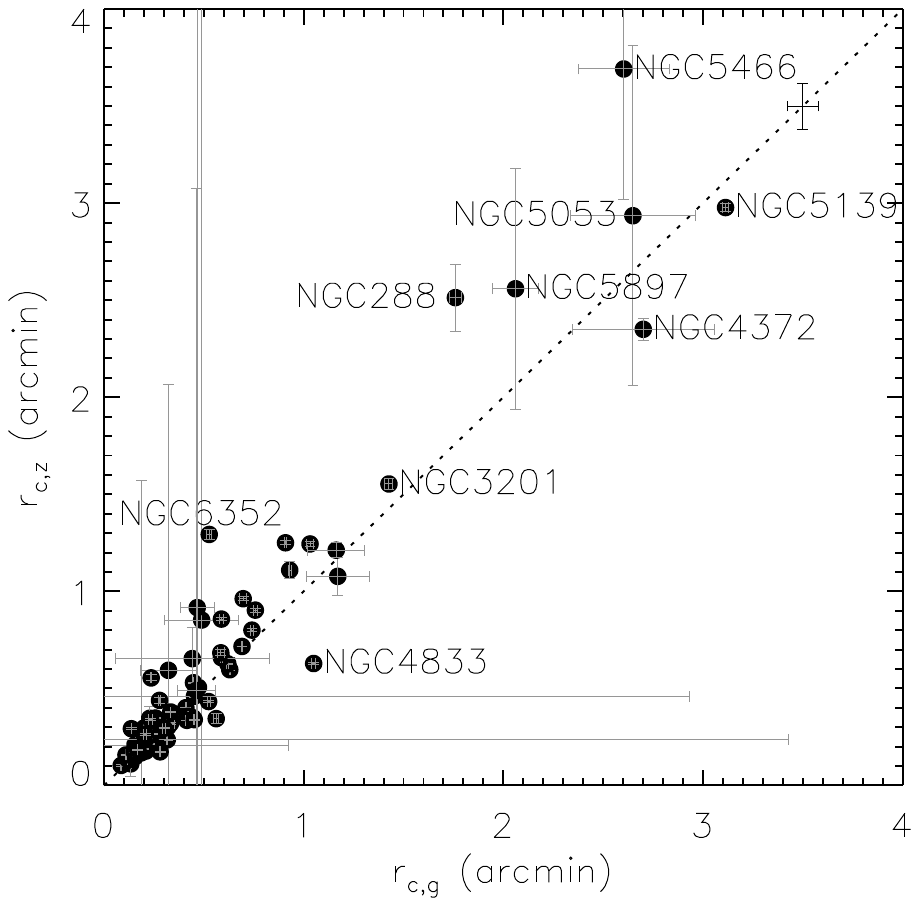}\includegraphics[scale=0.6,trim= 2.7cm 13.2cm 9cm 5.8cm]{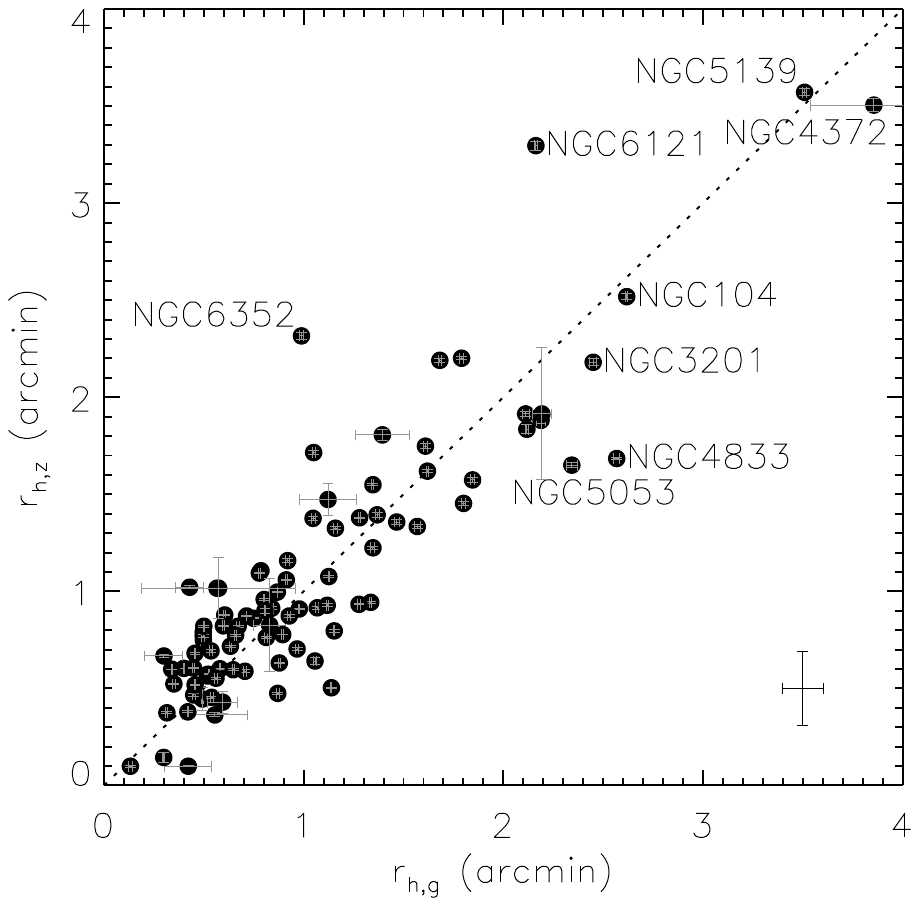}
\caption{Comparison of the King parameters derived based on observations with the $g$ and $z$ filters. The black error bars present the corresponding systematic errors, as listed in Table~\ref{tab:systerr}. } 
\label{fig:Kinggz_comp}
\end{figure*}

Fig.~\ref{fig:Kinggz_comp} compares the core radii and half-light radii based on the $g$ and $z$ filters. We do not find evidence for a 
systematic trend in the core radius or the half-light radius. 

GCs close to the Galactic plane are known to suffer from differential reddening \citep{AlonsoGarcia2011}. Observations taken at longer 
wavelengths are less affected by the extinction. Therefore, we expect a correlation between the differential reddening and the $gz$ structural 
parameter differences, if the differential reddening plays a significant role in the determination of the King parameters. However, we do not find 
any correlation and conclude that the differential reddening does not strongly bias our parameter determinations.

\section{Comparison with the literature}\label{sec:complit}

We now compare our results with previous determinations of structural parameters from the literature. First, we compare with the 2010 
edition of \cite{Harris1996}, which is largely a compilation of the \cite{McLaughlin2005} SB profile study. Then we also compare with the 
results of \cite{Miocchi2013}, who used star count density profiles to determine the structural parameters.

It is clear that some clusters have lower quality data, resulting in more uncertain parameters. We therefore restrict our analysis
to a smaller, ''clean'' sample where we can trust our determinations of the main structural parameters. We impose following constraints: 
\begin{itemize}
\item $r_c>\sigma(r_c)$: The core radius has to be significantly different from zero (based on the bootstrap errors, not including the systematic errors). 
\item $\mu_0<20 \text{\,mag}/\arcsec^2$: clusters fainter than this SB have very large photometric errors. SB points are often not significantly different from zero. 
\end{itemize}

These conditions are met for 12, (80, 41, 42, 87) in the $u$ ($g,r,i,z$, respectively)-band. We have experimented with SB criteria including the effects of foreground extinction (basically imposing magnitude limits on the apparent SBs, not corrected for extinction), but this did not strongly affect the selected clusters for the clean sample. Therefore, we do not introduce restrictions on the foreground extinction as this would automatically exclude all GCs close to the Galactic plane. However, some disk/bulge clusters were removed from the sample, either because of the strong stellar contamination or because of the low apparent SB points (often not significantly higher than the sky, partly due to the foreground extinction). Because the number of clusters is largest in the $gz$-bands, we will focus on these filters during the analysis and discussion.

\subsection{Comparison to parameters based on SB profiles}\label{sec:compSB}

\begin{figure*}
\centering 
\includegraphics[scale=0.72,trim= 2.5cm 13.cm 10.7cm 1.3cm]
{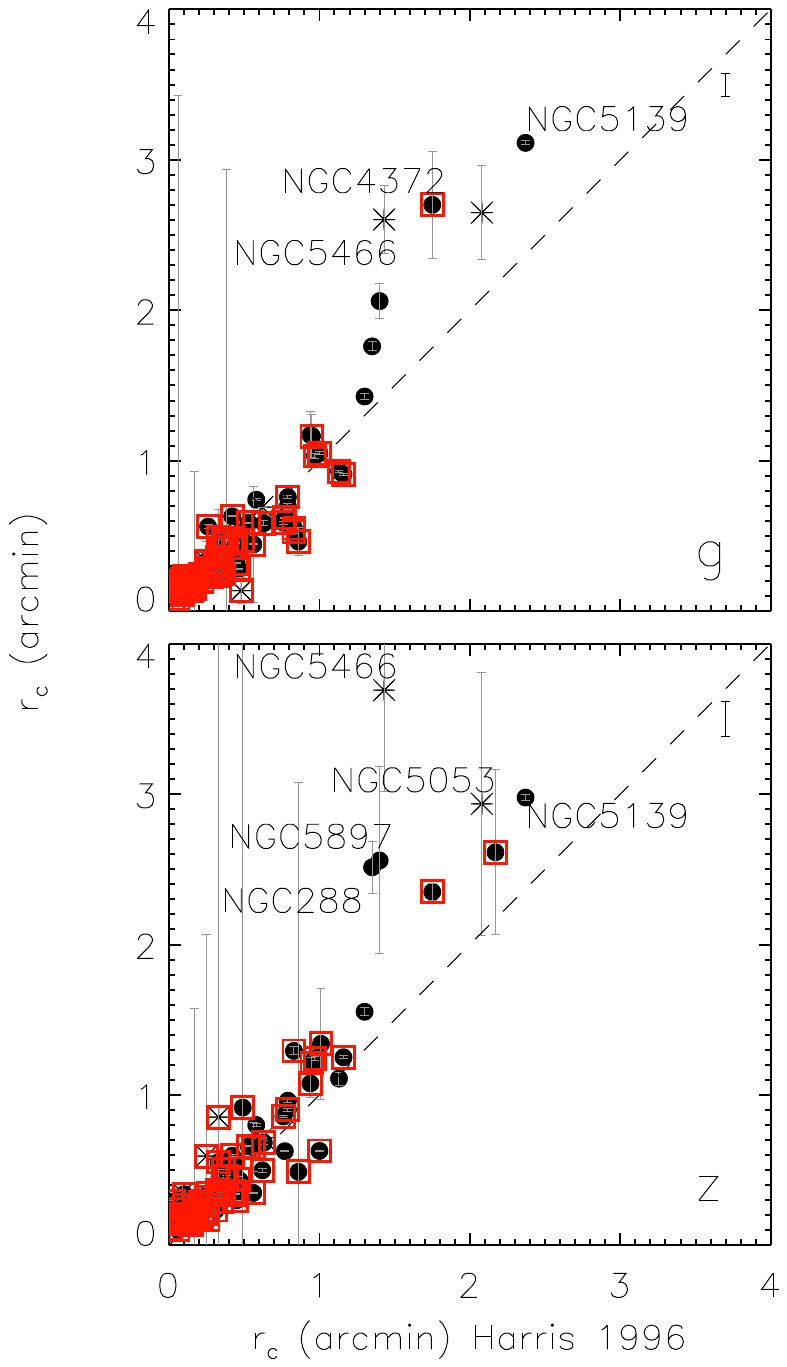}\includegraphics[scale=0.72,trim= 2.5cm 13.15cm 10.7cm 1.3cm]{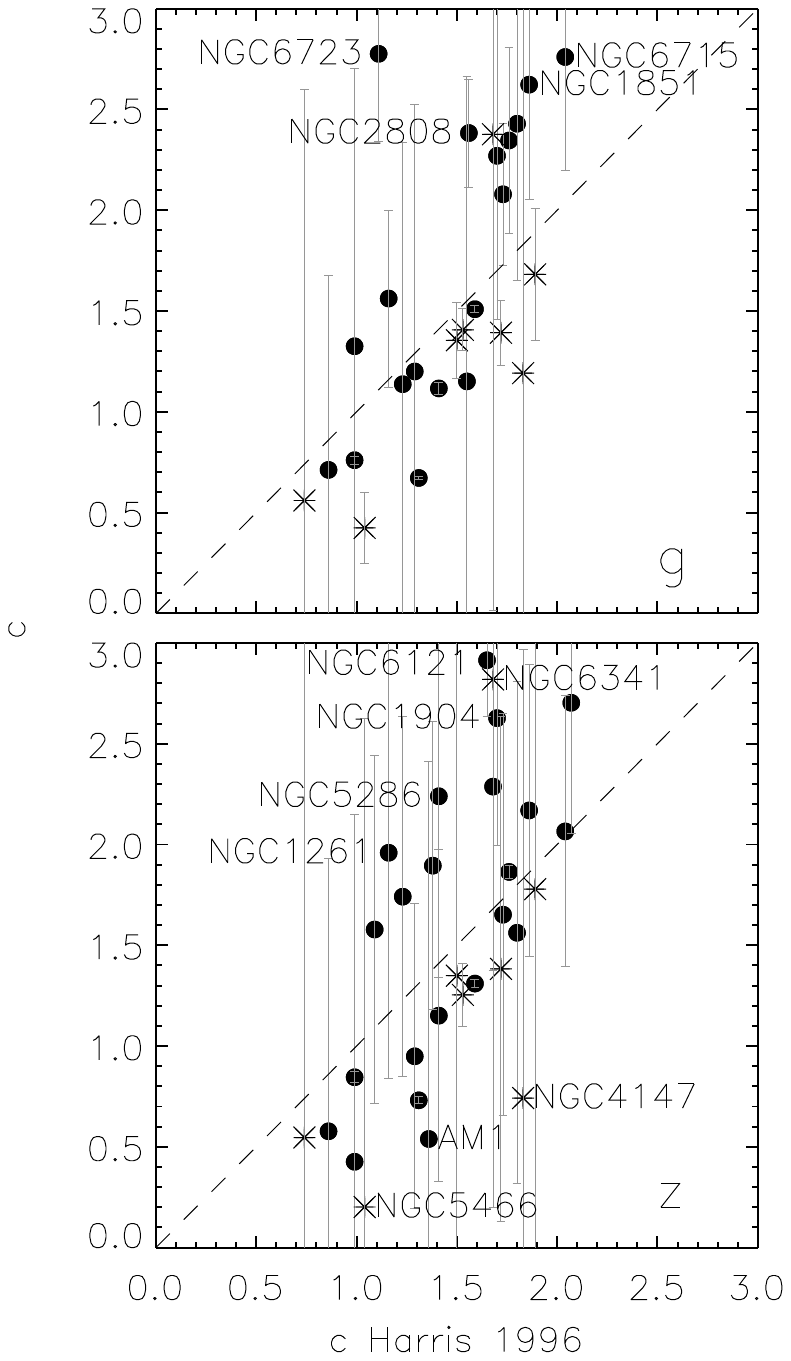}\includegraphics[scale=0.72,trim= 2.5cm 13.cm 10.7cm 1.3cm]{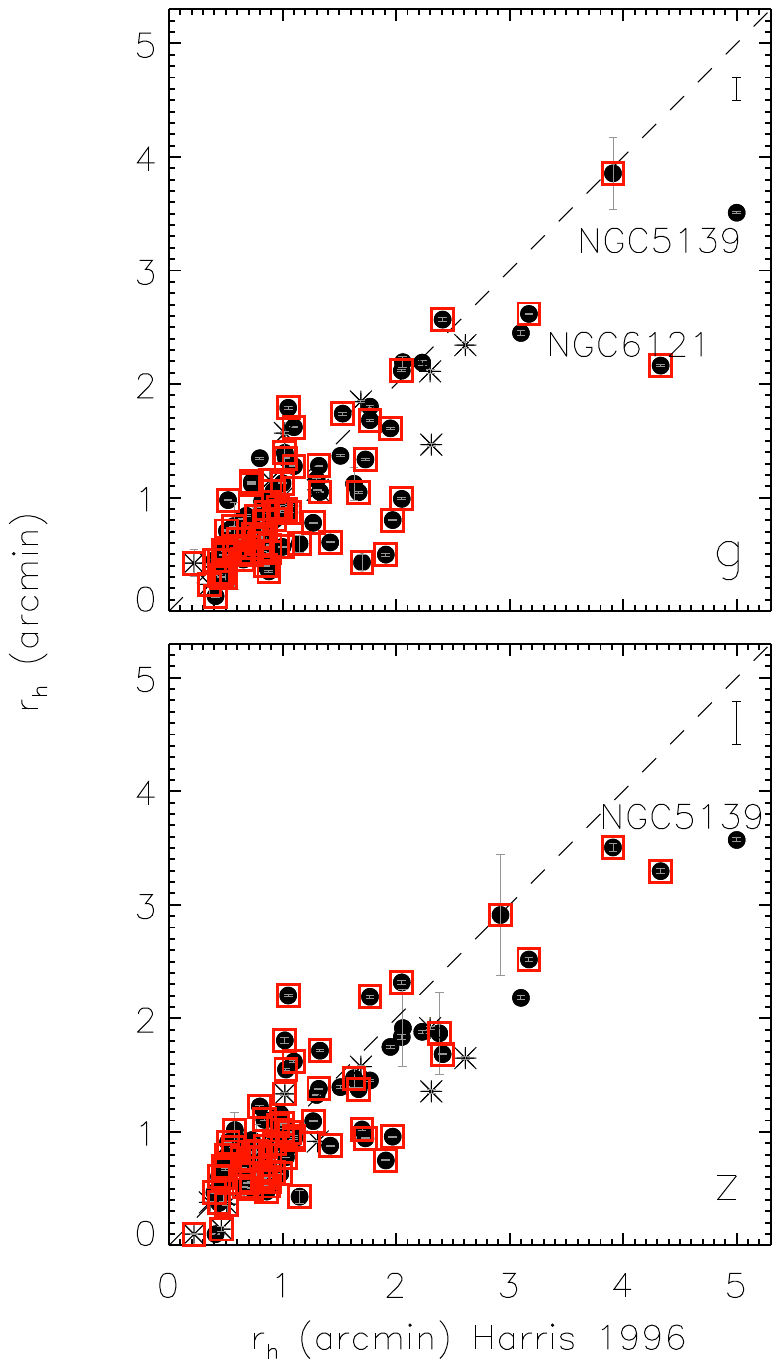}
\caption{Comparison of the structural parameters between our results (for the $g$- and $z$-band) and the literature \citep{Harris1996}. Filled circles are values based on CTIO SB profiles, supplemented with SDSS values represented by asterisks. The dashed line indicates the one-to-one correspondence. High-concentration clusters (with $c\ge2.5$) are indicated with red boxes in the left and right panels. Concentrations higher than $\sim 2.5$ are unreliable, which was also concluded from the simulations in Fig.~\ref{fig:mockc}. Therefore, clusters with $c>3$ are not included in the concentration panel. However, both $r_c$ and $r_h$ are recovered well for the bulk of the cluster sample. The systematic error (given in Table~\ref{tab:systerr}) is illustrated by the black error bar in the top-right corner of each panel. See text for more details. } 
\label{fig:compare_harris}
\end{figure*}

Fig.~\ref{fig:compare_harris} compares $r_c$, $c$ and $r_h$ to the literature \citep[][]{Harris1996}. The left panel of the figure compares 
the core radii (based on the $g$- and $z$-bands). The match for the well-populated clusters is very good, with the exception of NGC~5139. 
For this cluster, our fitting algorithm clipped the central SB point (for both $g$ and $z$ filters), which accounts for the larger value of the core radius. 
However, for poor clusters (indicated in the figure), our resulting $r_c$ can be biased. 
The middle panel compares our concentrations with \cite{Harris1996}. For the bulk of our sample, the concentrations are unreliable and will therefore not be discussed further. 

In the right panel of Fig.~\ref{fig:compare_harris} we show our half-light radii vs. the values in
the \cite{Harris1996} compilation. For smaller clusters ($r_h \lesssim 1\arcmin$), the data fall close to the
45-degree line, albeit with somewhat large scatter. For clusters subtending a larger angle on the sky, we
find that our $r_h$ are systematically lower, probably because of our poorer value of $r_t$ or a poor sky determination (see Section~\ref{sec:SBprofiles}). 

As an additional check, we selected a representative subsample of CTIO clusters and fitted S{\'e}rsic profiles in pixel space with \textsc{budda} \citep{Desouza2004,Gadotti2008}. Again, we found that the effective S{\'e}rsic radius (equivalent to the radius encompassing half of the cluster light) was significantly smaller (about a factor three) than the literature values. Both approaches suffer from the sky determination issues related to the limited FOV of the CTIO data. Nevertheless, the centre determination is incorporated in the \textsc{budda} fitting algorithm, hence we do not expect a strong bias due to central RGB stars. 

\begin{figure*}
\centering
\includegraphics[scale=0.6,trim= 2.8cm 13.2cm 9cm 6cm]{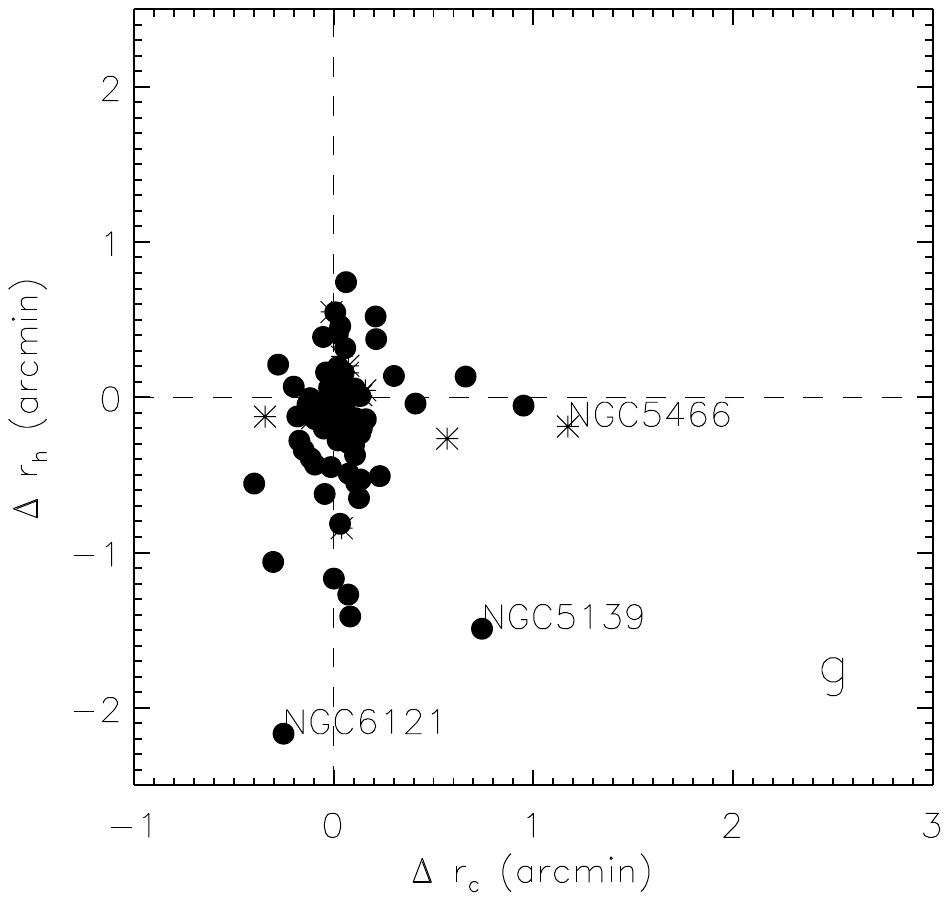} \includegraphics[scale=0.6,trim= 2.8cm 13.2cm 9cm 6cm]{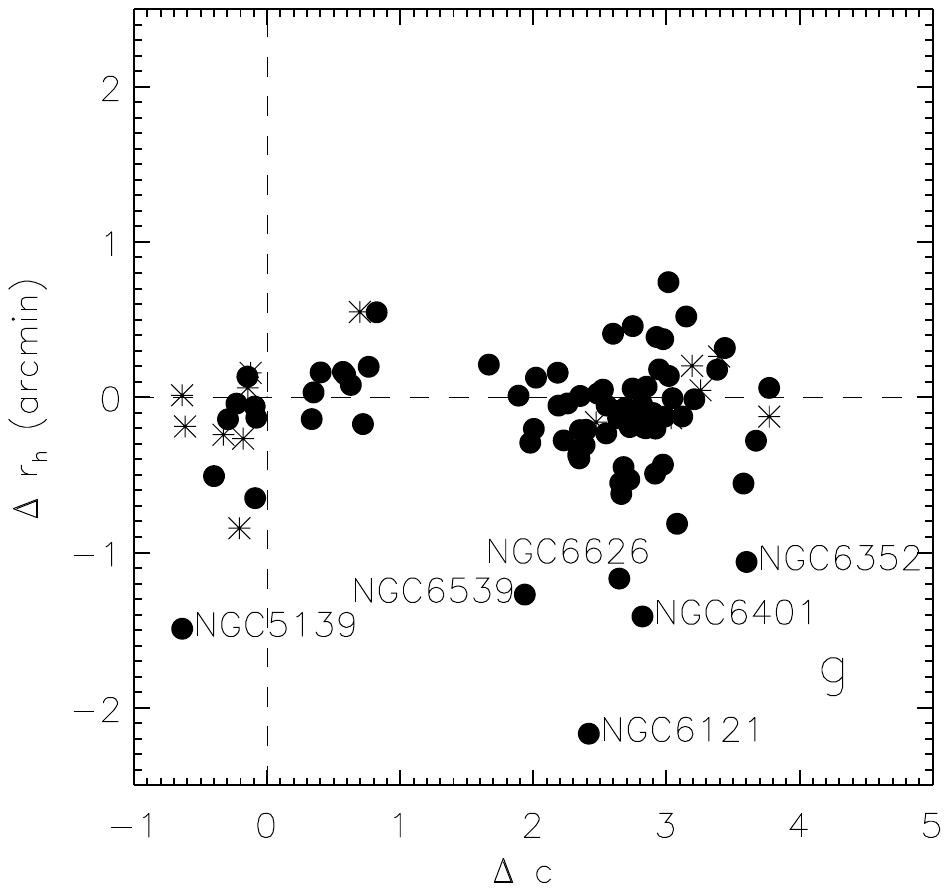}\includegraphics[scale=0.6,trim= 2.8cm 13.2cm 9cm 6cm]{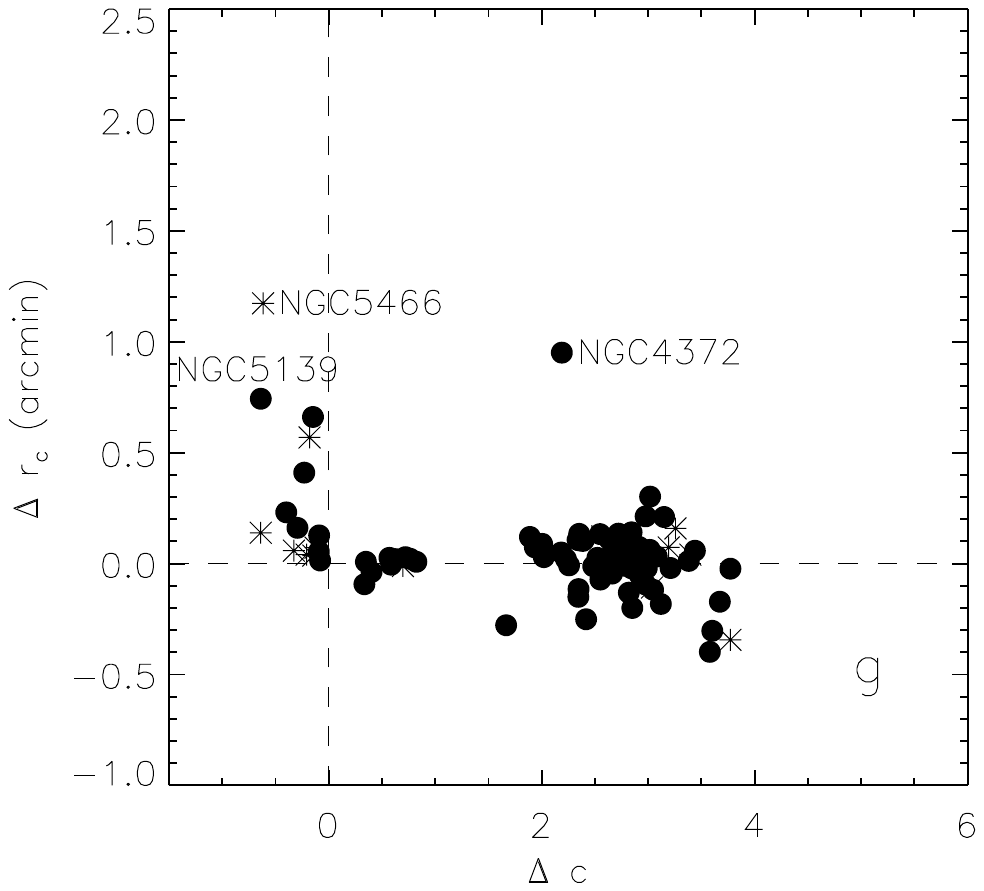}

\includegraphics[scale=0.6,trim= 2.8cm 13.2cm 9cm 5.6cm]{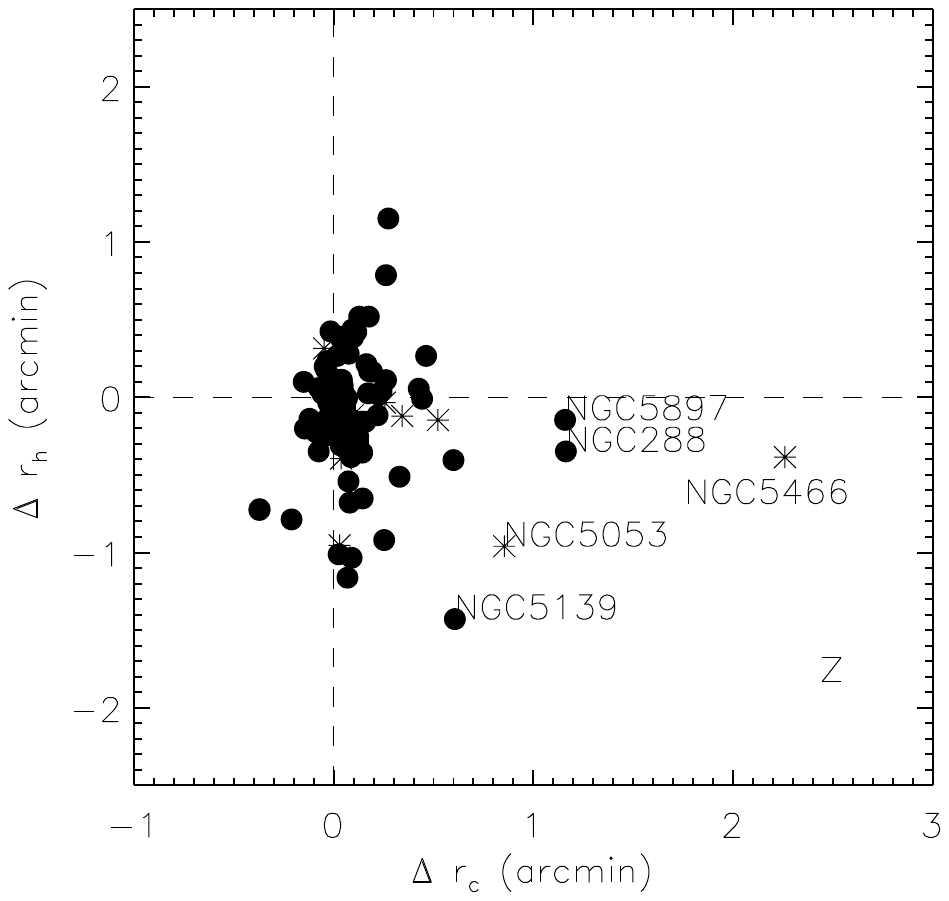} \includegraphics[scale=0.6,trim= 2.8cm 13.2cm 9cm 5.6cm]{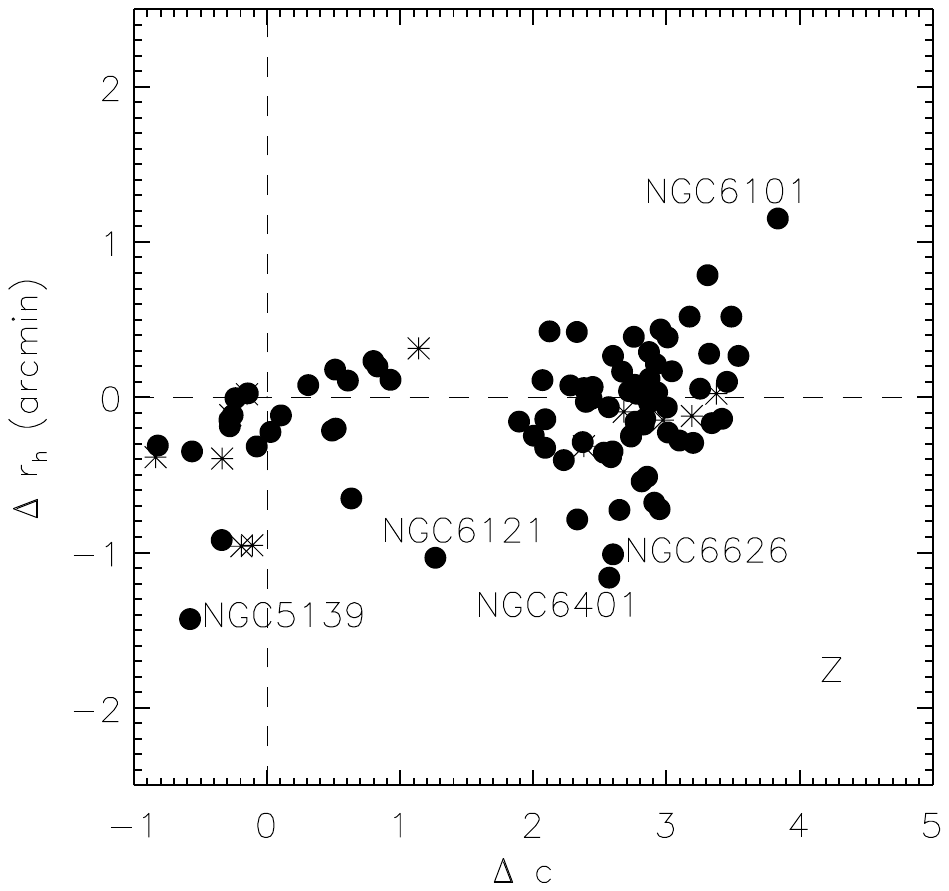}\includegraphics[scale=0.6,trim= 2.8cm 13.2cm 9cm 5.6cm]{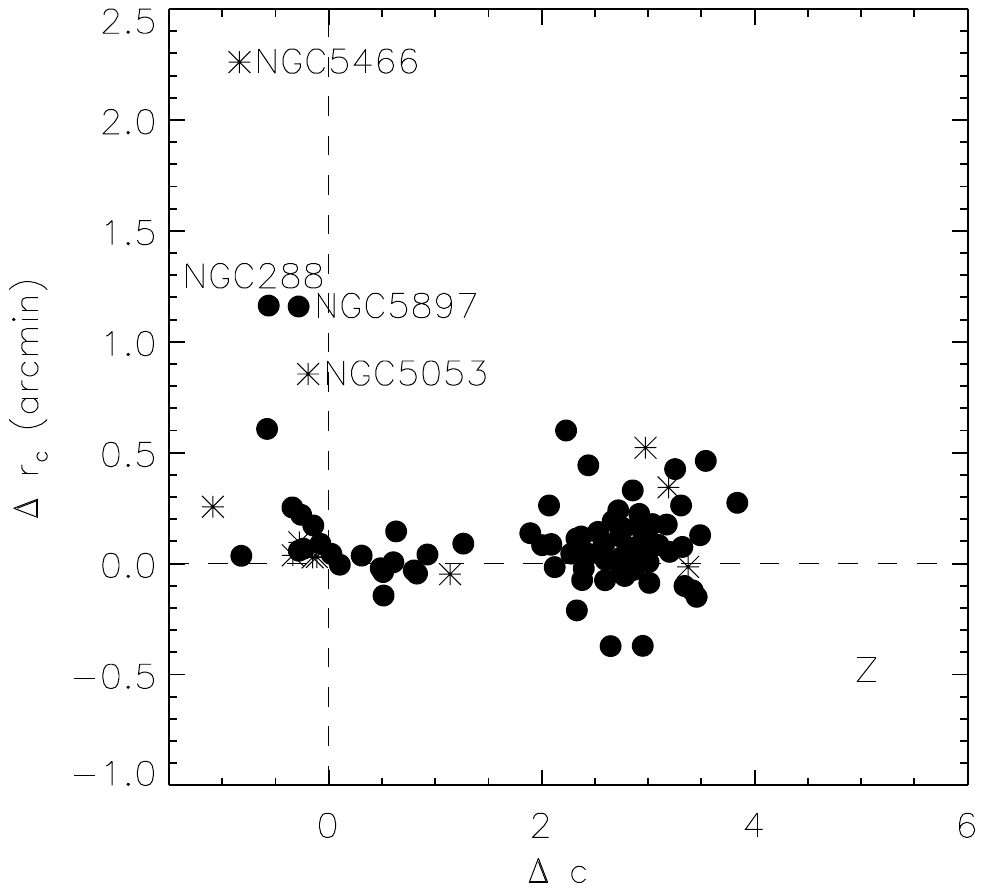}
\caption{Correlations between the differences of the King parameter determinations when compared to the literature ($\Delta p = p_{\text{this~work}}-p_{\text{Harris 1996}}$ for $p=r_c,c,r_h$).} 
\label{fig:deltarcrhc}
\end{figure*}

The King model parameters are fitted simultaneously. Therefore the discrepancies between the resulting parameters and the values given by \cite{Harris1996} could be correlated. In Fig.~\ref{fig:deltarcrhc} we study these correlations in more detail and conclude that no correlations between $\Delta r_c$, $\Delta r_h$ and $\Delta c$ are present. 

\subsection{Comparison to star count density profiles}\label{sec:compmiocchi}
Star count density profiles mainly trace the abundant main-sequence stars in GCs. SB profiles are vastly affected by bright evolved RGB stars, that sink to the cluster centre due to mass segregation \citep{Fregeau2002}. In contrast to SB profiles, star count profiles are not biased by the presence of sparse, bright stars \citep{Noyola2006}. Therefore, the latter are believed to be the most reliable way to determine structural parameters. 
However, our CTIO and SDSS data does not have the required resolution to derive such radial stellar density profiles. Based on a combination 
of high-resolution HST observations and ground-based observations, \cite{Miocchi2013} were able to construct star count profiles and derived 
structural parameters for 26 Galactic GCs. 

\begin{figure*}
\centering 
\includegraphics[scale=0.6,trim= 2.7cm 13.2cm 9cm 5.8cm]{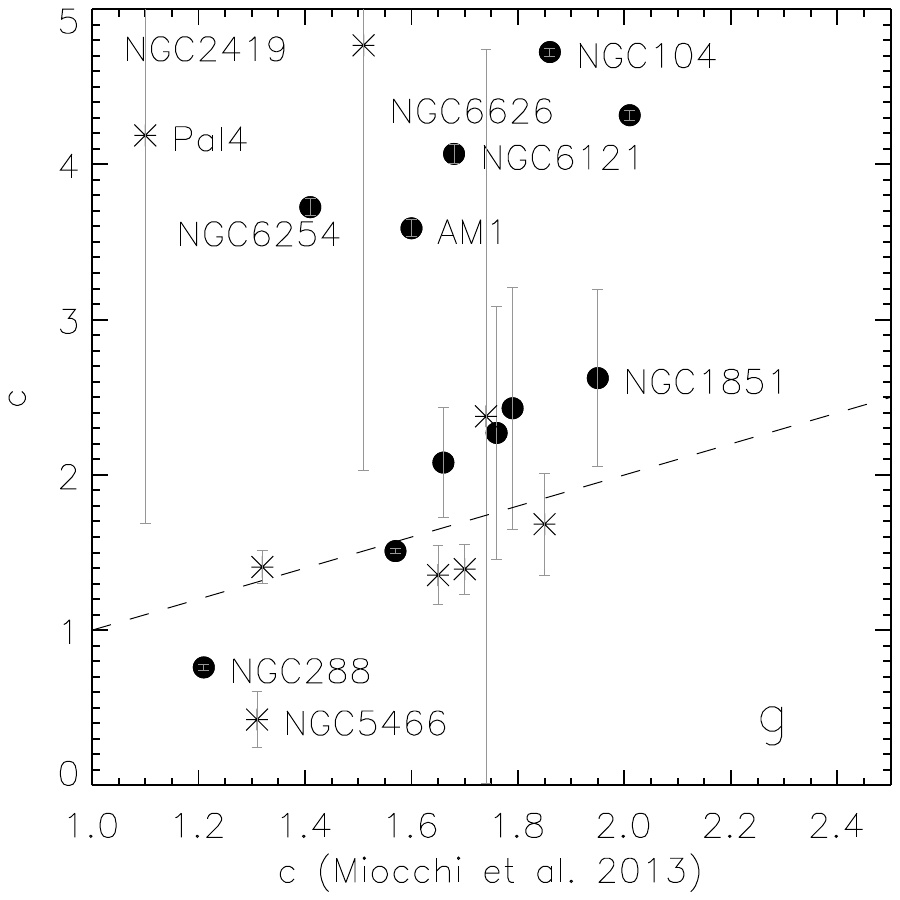}\includegraphics[scale=0.6,trim= 2.7cm 13.2cm 9cm 5.8cm]{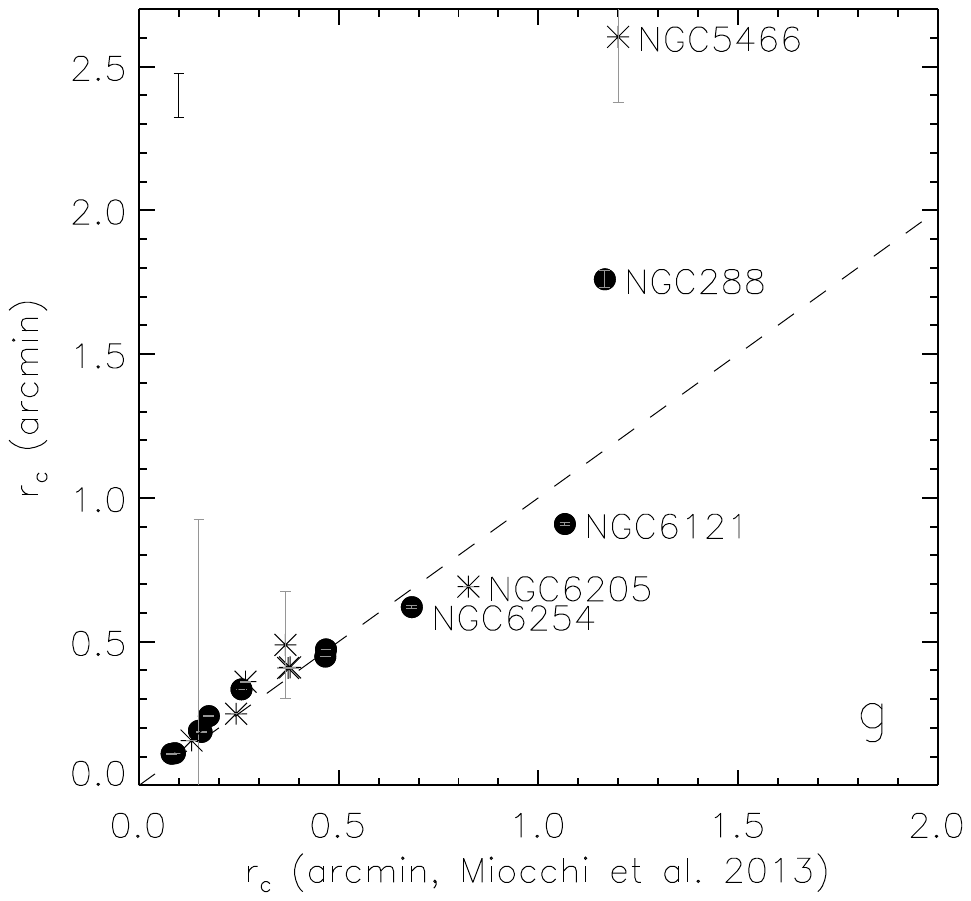}\includegraphics[scale=0.6,trim= 2.7cm 13.2cm 9cm 5.8cm]{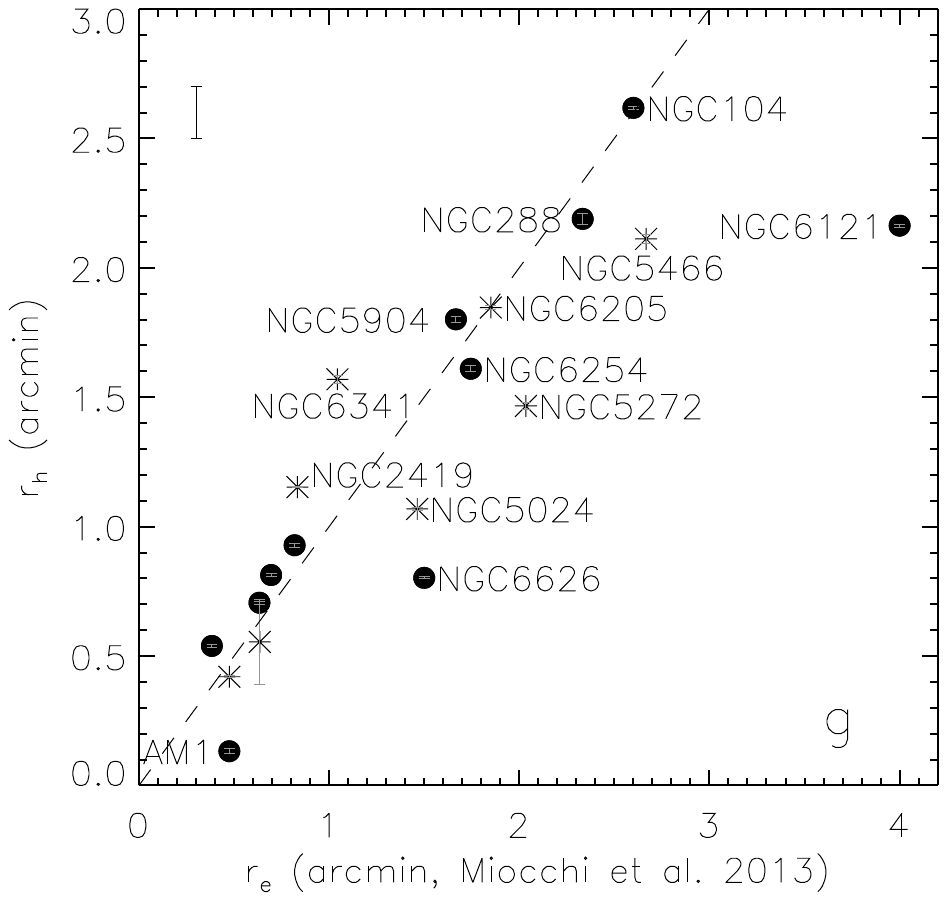}
\includegraphics[scale=0.6,trim= 2.7cm 13.2cm 9cm 5.5cm]{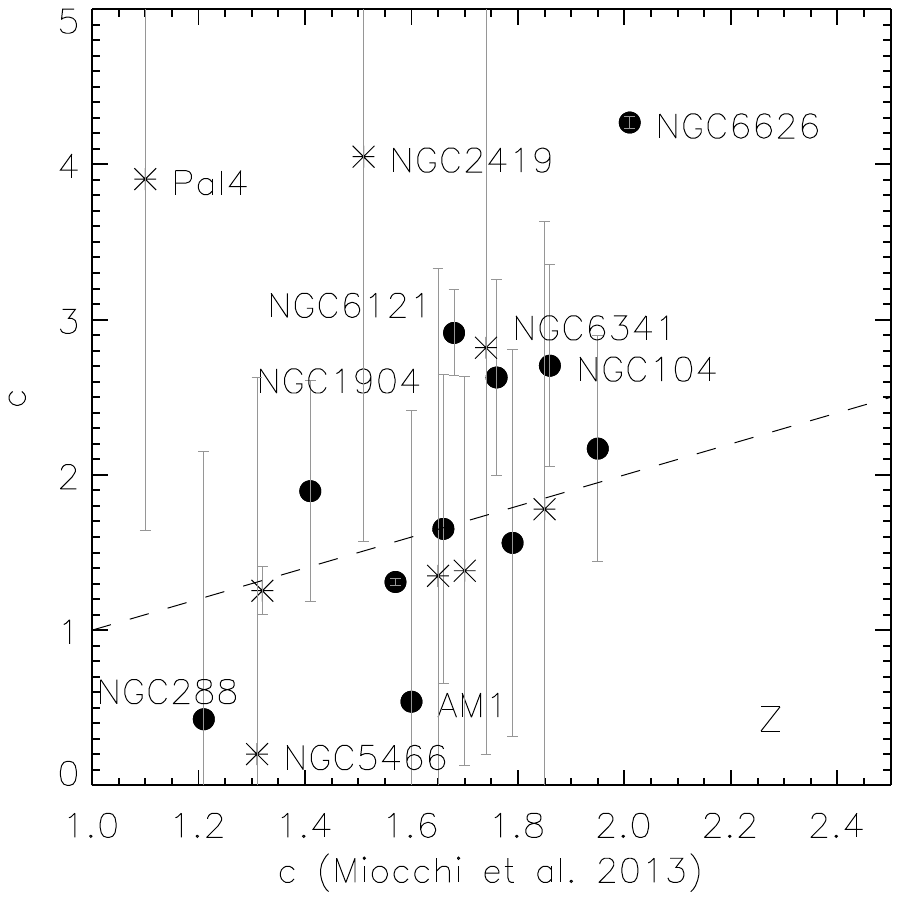}\includegraphics[scale=0.6,trim= 2.7cm 13.2cm 9cm 5.5cm]{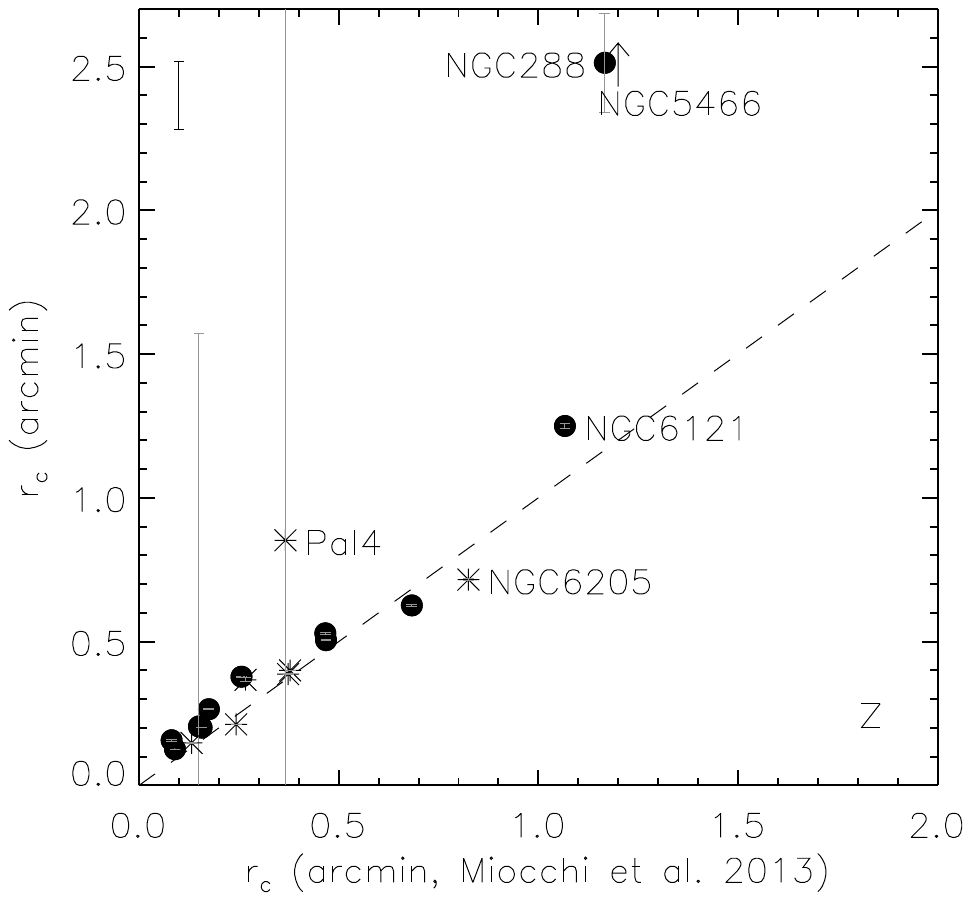}\includegraphics[scale=0.6,trim= 2.7cm 13.2cm 9cm 5.5cm]{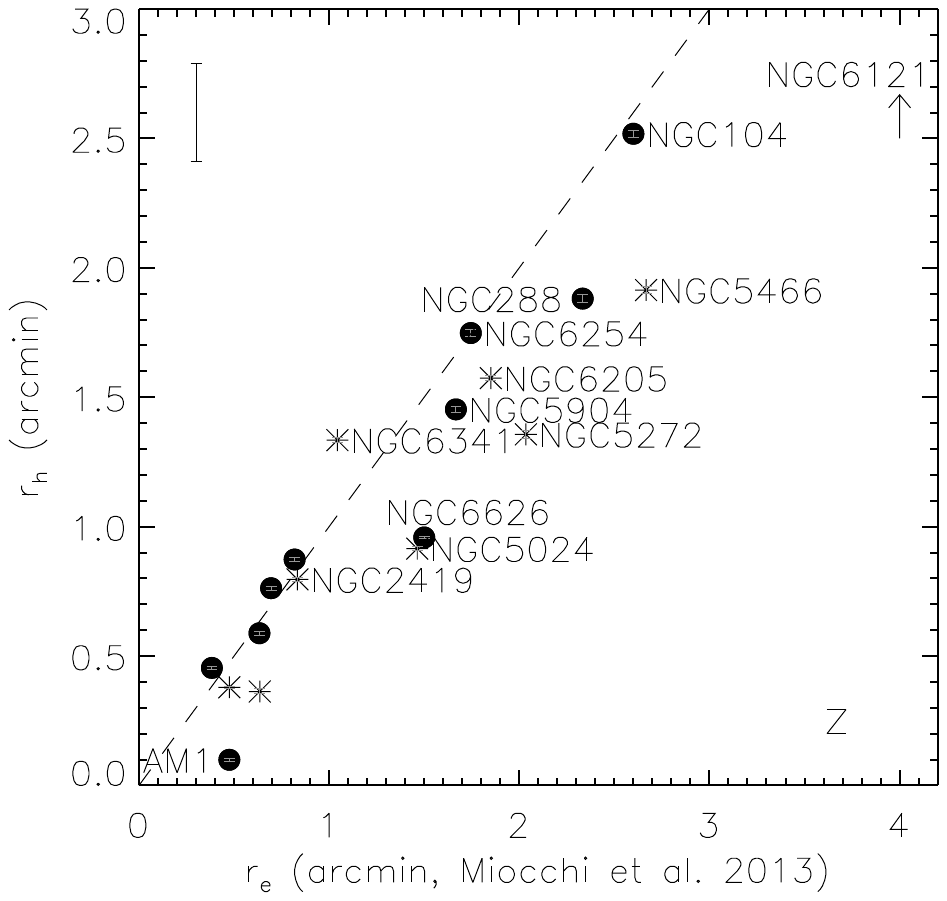}
\caption{Comparison between our King model parameters ($c$, $r_c$, $r_h$) and the parameters based on King model fits to star count density profiles \citep{Miocchi2013}. The dashed line indicates the one-to-one correspondence. Concentrations higher than $\sim 2.5$ are unreliable, which was also concluded based on the mock data in Fig.~\ref{fig:mockc} and from Fig.~\ref{fig:compare_harris}.The black error bar in the top-left corner of each panel illustrates the systematic error given in Table~\ref{tab:systerr}.}
\label{fig:Miocchi}
\end{figure*}

Fig.~\ref{fig:Miocchi} presents a comparison between our King model parameters and the parameters derived by \cite{Miocchi2013}. Our concentrations, shown in the left panels, again prove unreliable. The middle panels compare our newly derived core radii with the core radii derived by \cite{Miocchi2013} based on King model fits to star count density profiles. In general, these compare well, with some exceptions which are indicated in the figure. The right panels of Fig.~\ref{fig:Miocchi} show a comparison between our King model $r_h$ and the effective radii $r_e$ derived by \cite{Miocchi2013}. The scatter is rather large with some significant outliers.

\section{Discussion}\label{sec:discussion}

\subsection{Correlations between size and metallicity}\label{sec:size_met}

Several extragalactic studies found that red (metal-rich) GCs are systematically smaller (about $20\%$) than blue (metal-poor) GCs (see, e.g.,
\citealt{Kundu1998,Kundu1999,Puzia1999,Larsen2001}). \cite{Larsen2003} argue that the difference is originated by a projection effect, due 
to a correlation between GC size and Galactocentric distance, combined with different radial distributions of the metal-poor and metal-rich 
GC subpopulations (with the metal-rich GCs being more centrally located; see e.g. \citealt{KisslerPatig1997,Lee1998,Cote2001,Dirsch2003}). 
At larger distances from the galactic centre, the projection effect is less strong, hence the trend should disappear in the galactic outskirts. 
However, \cite{Harris2009} studied GCs from massive galaxies and found that the size differences between red and blue GCs remained at 
large distances. An alternative scenario was proposed by \cite{Jordan2004}, who advocate that the correlation between colour (metallicity)
and $r_h$ is a consequence of mass segregation and the longer lifetimes (for a given mass) of more metal-poor stars.

\begin{figure*}
\centering 
\includegraphics[scale=0.87,trim= 2.7cm 13.2cm 9cm 5.8cm]{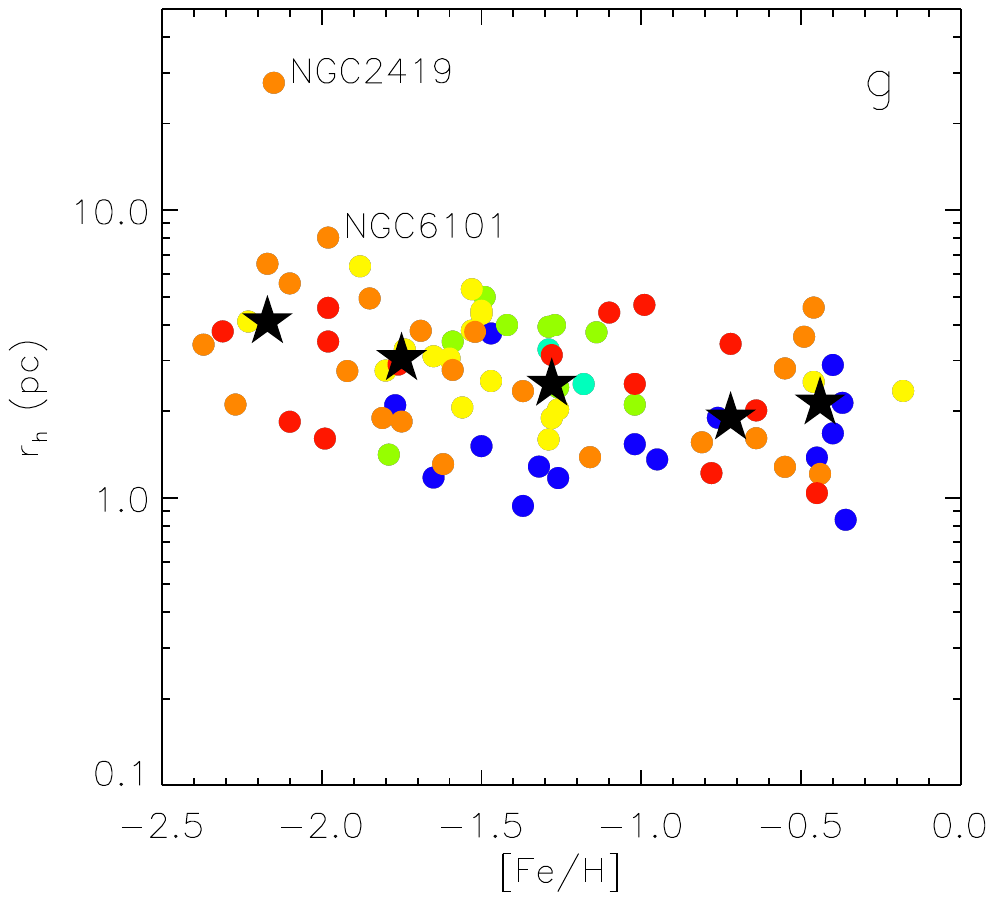}\includegraphics[scale=0.87,trim= 2.7cm 13.2cm 9cm 5.8cm]{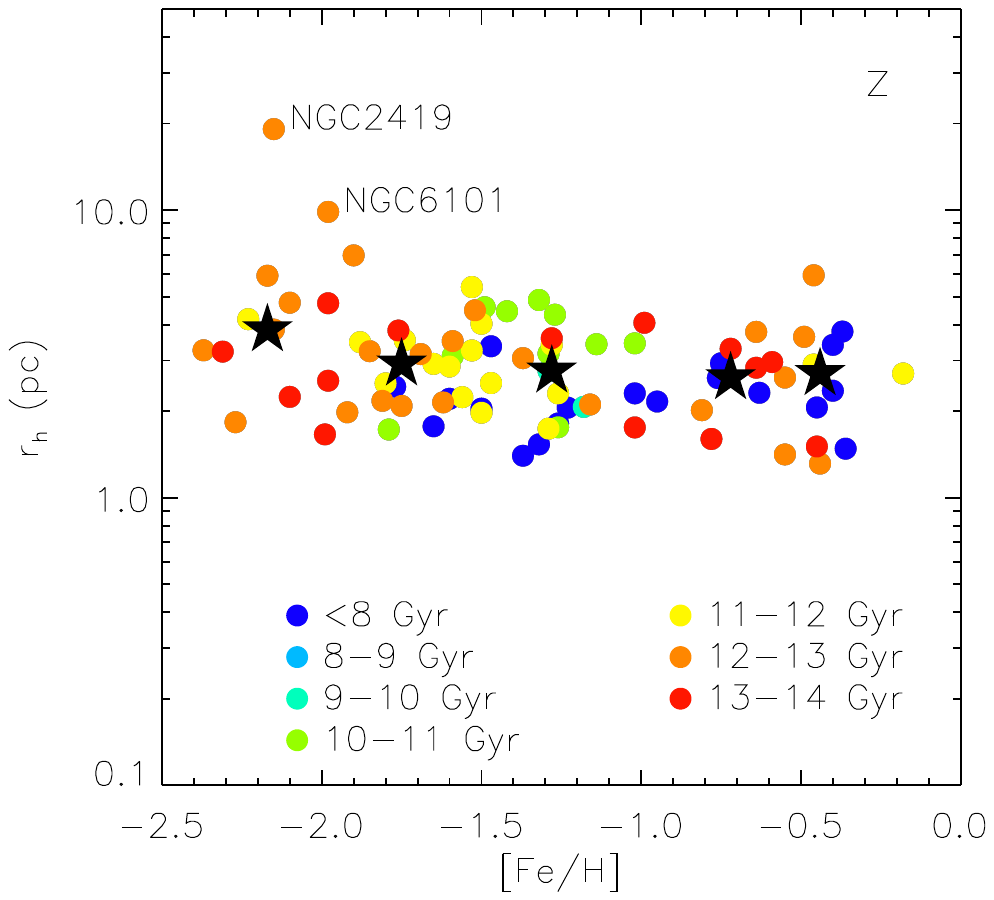}
\caption{Correlation between $r_h$ and [Fe/H]. The black filled stars present the median values for metallicity bins of 0.5 dex. } 
\label{fig:FeHrhpc}
\end{figure*}

In Fig.~\ref{fig:FeHrhpc} we show the half-light radii as a function of the metallicity \citep{Harris1996}, colour-coded by age (which are largely obtained from \citealt{Forbes2010}, with some other references as described in \citetalias{Vanderbeke2014b}). The correlation between both parameters is rather weak (Spearman's rank correlation coefficient is $\rho_{s,g}\sim-0.4$ with a significance level of $4.24\times10^{-4}$ for the $g$-band and $\rho_{s,z}\sim-0.23$ with a significance level of $3.12\times10^{-2}$ for the $z$-band). Nevertheless, \cite{Jordan2004} stress that large samples are needed to accurately 
determine the {\it average} behaviour. Although Galactic GCs are not so numerous, we still try to recover the average behaviour by showing the median [Fe/H] and $r_h$ in 0.5~dex metallicity bins. These median $r_h$ values largely agree with the trend predicted by \cite{Jordan2004}: $r_{h,g}\sim$4.1 pc, 3.1 pc, 2.5 pc, 1.9 pc and 2.1 pc for the different metallicity bins ($r_{h,z}\sim$3.8 pc, 2.9 pc, 2.8 pc, 2.6 pc and 2.7 pc), thus the $r_h$ difference between $g$ and $z$-band for the different metallicity bins is 0.3 pc, 0.2 pc, $-0.3$ pc, $-0.7$ pc and $-0.6$ pc. However, the actual size differences are much larger than predicted by \cite{Jordan2004}. The interpretation of the trend between cluster size and iron abundance is not that straightforward. Therefore, we will revisit this issue in Section~\ref{sec:Galdist}, where we discuss the correlation between the GC size and the Galactocentric distance.

Note that for the $g$-band, the youngest (i.e. $<8$\,Gyr) clusters are systematically below the median values for the corresponding metallicity bin, while this is less clear for the $z$-band. 

\subsection{Correlations between age and core radius}
\cite{Mackey2008} found a trend between age and core radii for the GCs in the Small and Large Magellanic Clouds, with the young  clusters (with ages between 10 Myr and 1 Gyr) showing a much smaller $r_c$ spread than the old GCs (older than 1 Gyr). These authors investigated two physical processes leading to large-scale core expansion: mass-loss due to rapid stellar evolution in a primordially mass-segregated cluster and dynamical heating due to stellar mass black holes. 

\begin{figure}
\centering 
\includegraphics[scale=0.87,trim= 2.7cm 13.2cm 9cm 5.8cm]{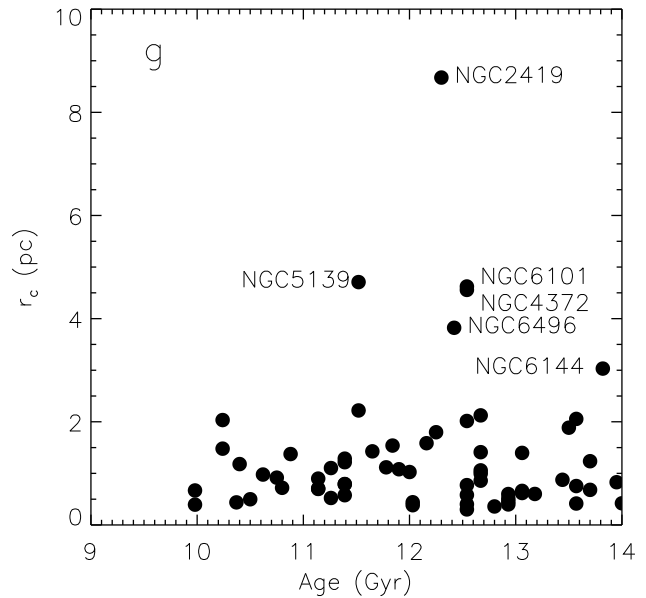}
\caption{Correlation between age and $r_c$. }
\label{fig:agerc}
\end{figure}

Galactic GCs are preferentially old \citep{Forbes2010}, hence one would expect to see a large range of $r_c$. However, Fig.~\ref{fig:agerc} 
shows that the majority of the clusters have small $g$-band core radii. \cite{Mackey2008} believe this is a consequence of the positions 
of the GCs in the Galaxy: because a large fraction of the MW GCs reside in the inner Galaxy, tidal forces are expected to rapidly destroy loosely bound clusters. The same conclusions can be drawn based on the $z$-band core radii. Both NGC~2419 and NGC~5139 (Omega Cen) have properties that hint to an extragalactic origin, which might explain their offset to the general trend. 

\subsection{Correlations with the Galactocentric distance}\label{sec:Galdist}

\begin{figure*}
\centering 
\includegraphics[scale=0.5,trim= 2.5cm 13.2cm 1cm 1cm]{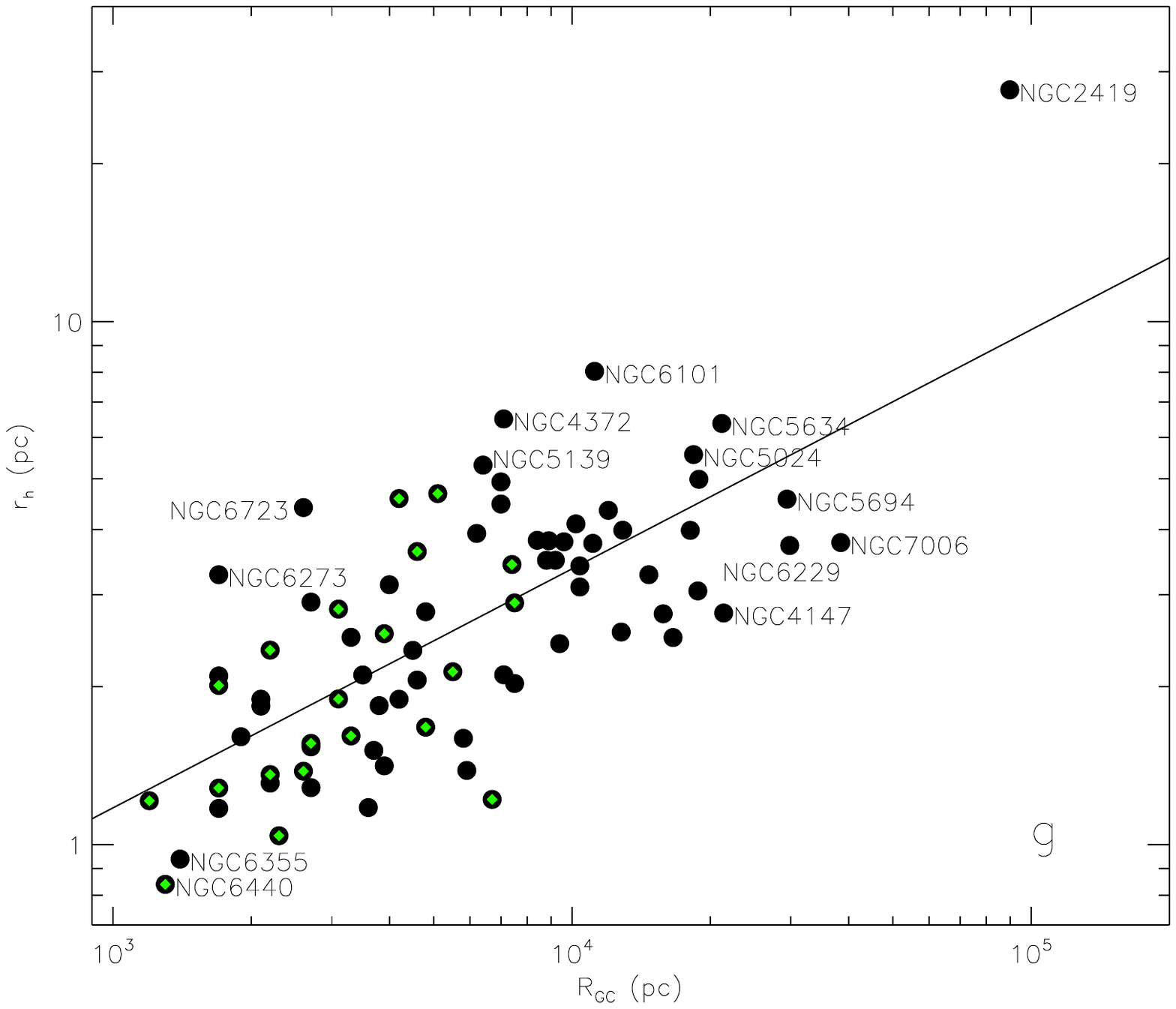}\includegraphics[scale=0.5,trim= 2.5cm 13.2cm 1cm 1cm]{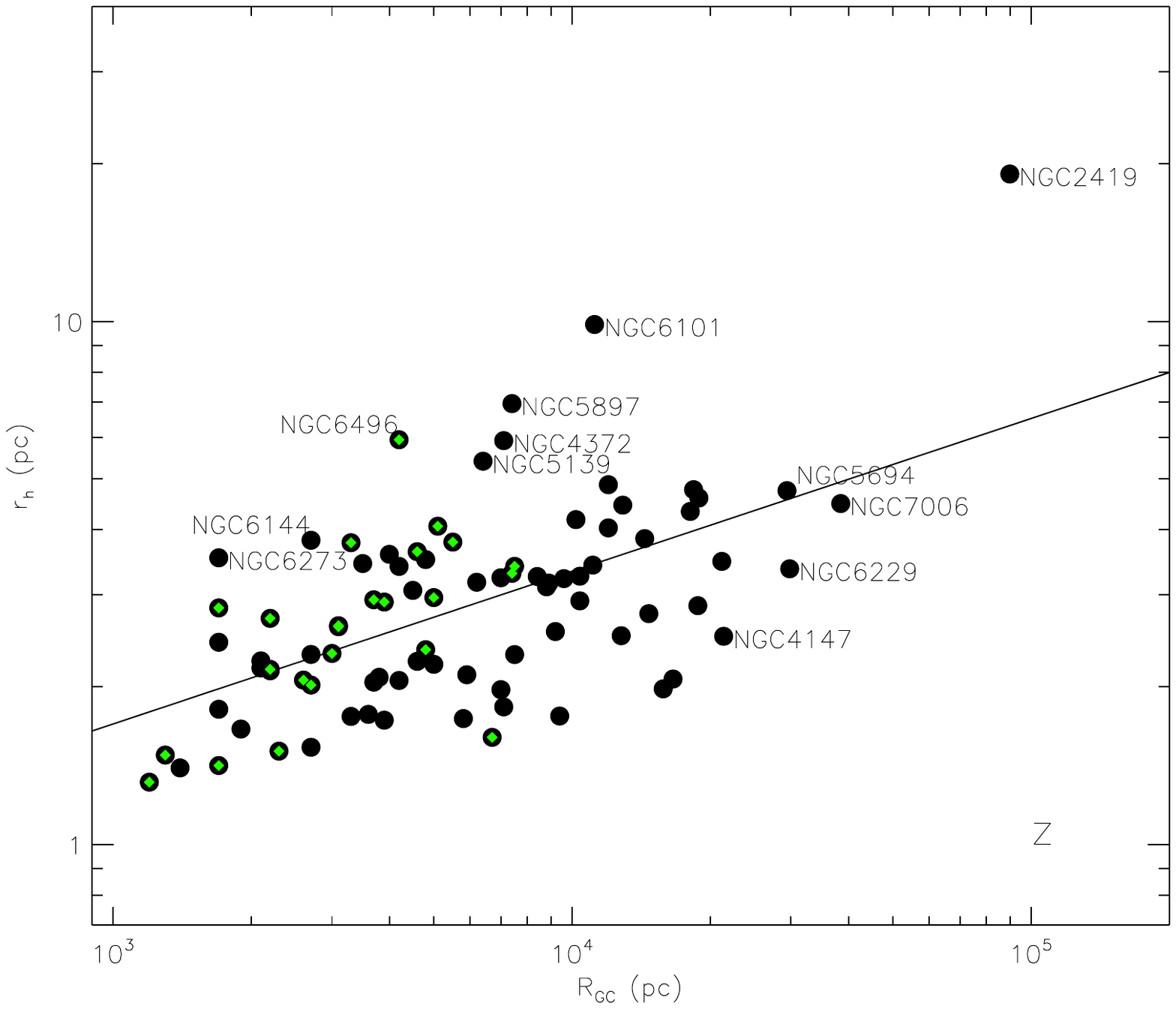}
\caption{Correlation between the half-light radius $r_h$ and the Galactocentric distance $R_{\text{GC}}$ for the $g$- and $z$-band. Small green filled diamonds indicate the GCs with $[Fe/H]>-1$. The solid lines present robust fits to the data points of the clean sample (given by Eqs.~\ref{eq:rhrgc_g} and \ref{eq:rhrgc_z}). }
\label{fig:rhvsrgc}
\end{figure*}

\cite{Vandenbergh1991} and \cite{McLaughlin2000} have proposed the existence of a relation between the Galactocentric distance ($R_{GC}$) 
and the effective radius, arising from the tidal truncation (but see also \cite{Kundu1999}). 
Recently \cite{Ernst2013} provide a more theoretical approach on the implications of the correlation between the half-mass and the 
galactocentric radius. \cite{Miocchi2013} recover this correlation and show that it does not depend on other cluster properties, confirming its
likely dynamical origin. \cite{Puzia2014} studied the GC system of NGC~1399 and pointed at the influence of the GC orbit distribution on the 
evolution of the structural parameters. Fig.~\ref{fig:rhvsrgc} shows the observed correlation between the half-light radius vs. Galactocentric 
distance in our $g$- and $z$-band data. Based on a robust fit to the data points, we find:
\begin{equation}\label{eq:rhrgc_g}
\log r_h=-1.30\pm0.10+(0.45\pm0.05)\times \log R_{\text{GC}}  
\end{equation} 
for the $g$-band (Spearmans rank correlation coefficient $\rho_{s,g}$ is about 0.68 with a significance level of $2.44\times10^{-12}$), which is in good agreement with the empirical power law found by \cite{Vandenbergh1991} and \cite{Mackey2005}, and, in reasonable agreement with the scaling relation found by \cite{Miocchi2013}. The latter study, though limited in sample size, covers several GCs at large $R_{GC}$. 

Surprisingly, for the $z$-band, we find:
\begin{equation}\label{eq:rhrgc_z}
\log r_h=-0.65\pm0.09+(0.29\pm0.04)\times \log R_{\text{GC}},
\end{equation}
which is significantly different ($\rho_{s,z}\sim0.53$, significance level $\sim1.49\times 10^{-7}$). To further investigate the origin of the offset, 
we made similar robust fits to the clean $r$ and $i$-band samples. We obtain:
\begin{equation}\label{eq:rhrgc_r_subsample}
\log r_h=-0.98\pm0.13+(0.38\pm0.07)\times \log R_{\text{GC}}  
\end{equation}
for the $r$-band ($\rho_{s,r}\sim0.57$, significance level $7.71\times10^{-5}$) and 
\begin{equation}\label{eq:rhrgc_i_subsample}
\log r_h=-0.76\pm0.12+(0.33\pm0.06)\times \log R_{\text{GC}}  
\end{equation}
for the $i$-band ($\rho_{s,i}\sim0.60$, significance level $2.69\times10^{-5}$), hence we find evidence for a decreasing slope for redder bandpasses. 
Clusters are redder outwards, either because of tidal truncation or because shocks with the Galactic centre environment speed up mass segregation. Colour gradients will be discussed in full detail in Section~\ref{sec:colgrad}. It will become clear that the colour gradients are mainly caused by the influence of RGB stars and are not representative for the cluster count densities.

\cite{Georgiev2014} find that nuclear star clusters (with sizes ranging between GC and UCD sizes) have smaller effective radii 
when measured in bluer filters. Now we will study the size differences of GCs when measured in different filters, and, we will try to link those differences to the discrepancy in the $R_{GC}-r_h$ relations. 

In Section~\ref{sec:size_met} we found tentative evidence for the existence of a median size difference related to the cluster metallicity: metal-rich clusters, which are preferentially located close to the Galactic centre, have larger $r_h$ when measured in redder filters. Here we define $\Delta r_{h,g-z}$ as the difference (in arc minutes) between the $g$-band and $z$-band half-light radii. For the entire sample, the median (mean) of $\Delta r_{h,g-z}$ is about $-0.06\arcmin$ ($-0.07\arcmin$, respectively), which is consistent with a zero difference within the errors. However, for GCs with $R_{GC}$ lower than 5 kpc, the median (mean) of $\Delta r_{h,g-z}$ is about $-0.19\arcmin$ ($-0.23\arcmin$, respectively), while for GCs at larger distances to the Galactic centre, the median (mean) of $\Delta r_{h,g-z}$ is about $0.09\arcmin$ ($0.08\arcmin$, respectively). Therefore, close to the Galactic centre, clusters have redder outskirts (or, equivalently, bluer centres). For GCs with $E(B-V)<0.1$ ($E(B-V)\ge0.1$), the median of $\Delta r_{h,g-z}$ is about $0.08\arcmin$ ($-0.15\arcmin$, respectively), hence GCs at low reddening appear larger in $g$ than in $z$, while GCs at high reddening appear smaller in $g$ than in $z$. Similar correlations are found when using the absolute distance above the Galactic plane instead of the extinction value.

To further scrutinise the effects of metallicity and tidal disruption on the cluster size, we make separate fits for the metal-rich and metal-poor subsamples of the clean sample. 
 Because the metal-rich GCs do not span a wide range in Galactrocentric distances, we fix the slope of the relation to the value found in Eqs.~\ref{eq:rhrgc_g} and \ref{eq:rhrgc_z} and find that the intercept for metal-rich and metal-poor subsamples is not significantly different. This points to a scenario in which the origin of the size difference is related to the Galactocentric distance, and, not to [Fe/H] (hence in contrast to the tentative evidence found in Fig.~\ref{sec:size_met}). 
 
\cite{Puzia2014} found a flattening of the $R_{GC}-r_h$ correlation in NGC~1399 beyond $\sim20$\,kpc, while \cite{Miocchi2013} did not find evidence for a flattening of the latter relation in the outskirts of the Milky Way. Some GCs at $\sim20$\,kpc with good fits are located below the general trend and could point at a flattening of the correlation. Nevertheless, the scatter on the correlation is significant.

To detect correlations between the deviation to the general trend and other parameters, we define 
\begin{equation}\label{eq:Drgcrh}
D_g=-1.30+0.45\times \log(R_{\text{GC}})-\log(r_h)
\end{equation}
for the $g$-band (and a similar definition for the $z$-band; see \citealt{Vandenbergh2012} for a similar approach). We searched for correlations between this new parameter and age, HB index, absolute magnitude and $[Fe/H]$, but did not find any significant trends and 
confirm therefore the purely dynamical origin of the correlation between $r_h$ and $R_{GC}$, concurring with the recent literature \citep{Vandenbergh2012,Miocchi2013}. We also looked for correlations between the absolute distance $D$ and the absolute distance to the Galactic plane, which could reveal some insights on the influence of tidal shocks during disk passages on the GC size. However, we did not find any clear trends.

\begin{figure*}
\centering 
\includegraphics[scale=0.75,trim= 2.7cm 13.2cm 9cm 5.8cm]{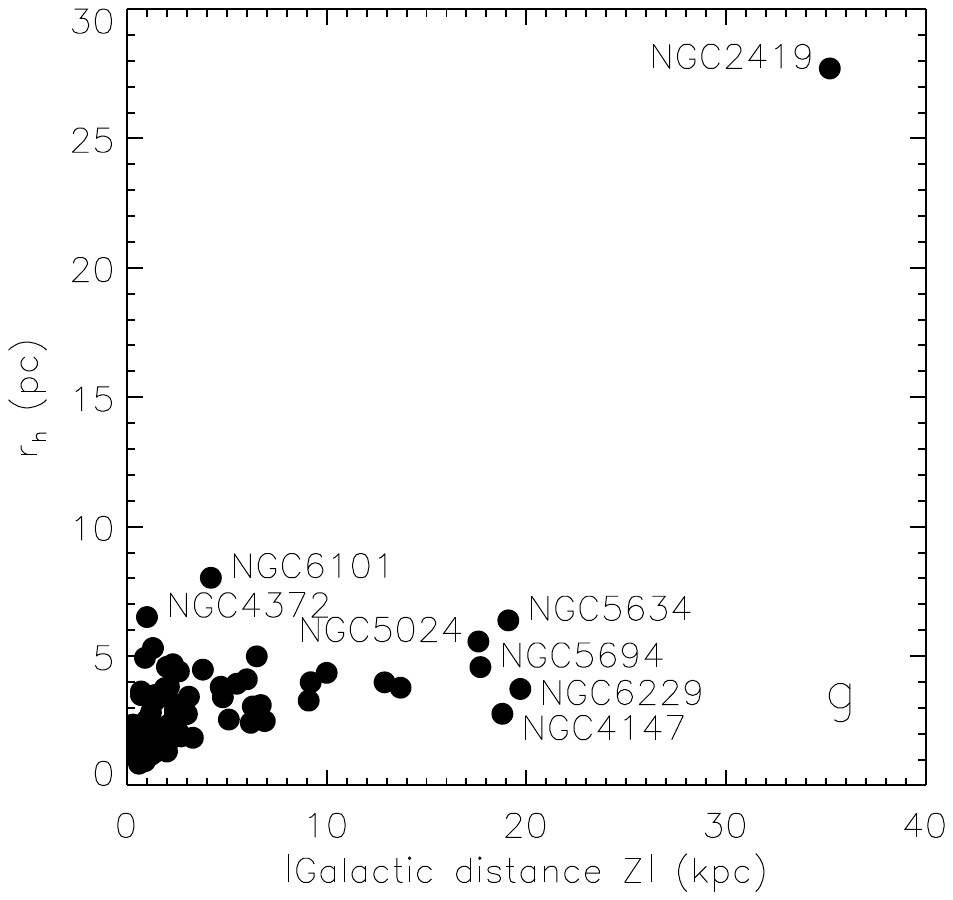}\includegraphics[scale=0.75,trim= 2.7cm 13.2cm 9cm 5.8cm]{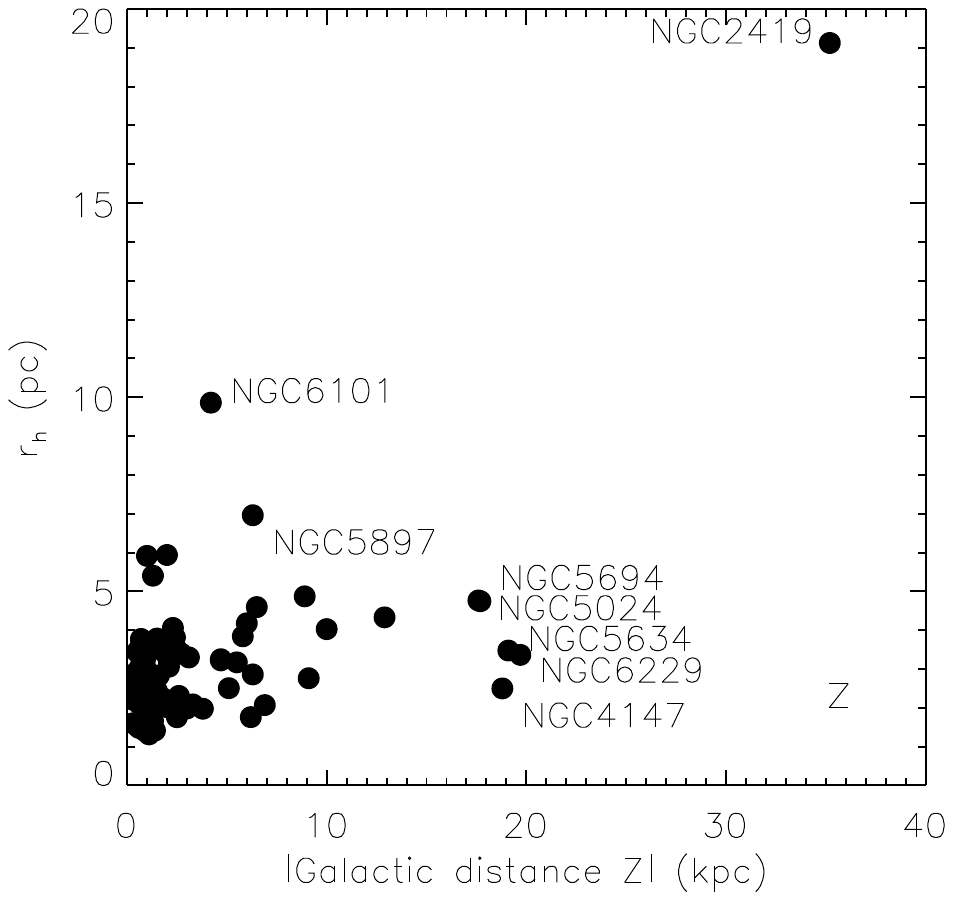}
\caption{Correlation between the absolute distance above the Galactic plane and $r_h$. }
\label{fig:absZ_rh}
\end{figure*}

Fig.~\ref{fig:absZ_rh} shows the correlation between the absolute distance above the Galactic plane and the half-light radii. The clusters located closer to the disk are generally smaller than those well above the Galactic plane, which can be ascribed to tidal stripping. For the $g$-band, the Spearmans rank correlation coefficient $\rho_{s,g}$ is about 0.61 with a significance level of $2.66\times10^{-9}$. 
For the $z$-band, the correlation is less strong: for the entire sample we find $\rho_{s,z}\sim0.46$ with a significance level of $8.66\times10^{-6}$.

\subsection{Distributions of the structural parameters} 
\begin{figure}  
\centering 
\includegraphics[scale=0.87,trim= 2.7cm 13.2cm 9cm 5.8cm]{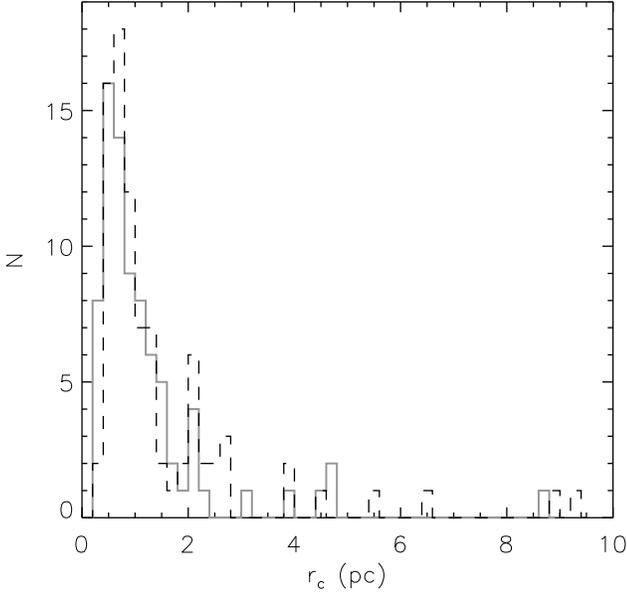}
\caption{The $g$-band (grey solid line) and $z$-band (black dashed line) distributions of the core radii.  }
\label{fig:histrc}
\end{figure}

The structural parameters of a cluster can be taken as a broad measure of its dynamical evolution. In \cite{Trager1995} the distribution of
core radii of globular clusters was found to be bimodal, with about 20\% of clusters showing central light excesses consistent with core collapse.
Cusps and similar excess light features appear to be relatively common in globular clusters in our Galaxy \citep{Vesperini2010} and the Large 
Magellanic Cloud \citep{Mackey2003}. In Fig.~\ref{fig:histrc} we show the $gz$ $r_c$ distributions. We do not find clear signs for bimodality, in contrast to \cite{Trager1995}. 

\begin{figure}    
\centering 
\includegraphics[scale=0.87,trim= 2.7cm 13.2cm 9cm 5.8cm]{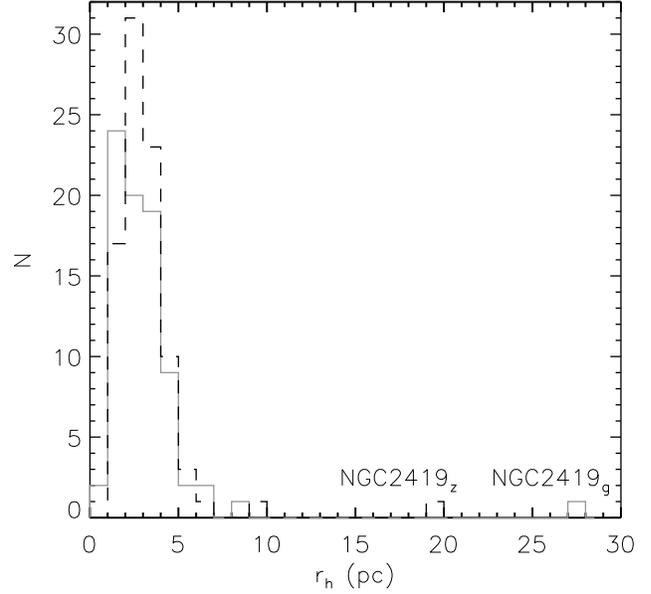}
\caption{The $gz$ distributions of the half-light radii. Legend as in Fig.~\ref{fig:histrc}. } \label{fig:histrh}
\end{figure}

Fig.~\ref{fig:histrh} presents the $gz$ distributions of the half-light radii. We find a median $r_h$ of 2.7 pc (2.9 pc) for the $g$-band
($z$-band, respectively), which is in good agreement with $r_h\sim3\text{\,pc}$ found for extragalactic GCs in the Virgo and Fornax 
clusters of galaxies (\citealt{Jordan2005,Masters2010}; see also e.g. Fig. 16 of \citealt{Puzia2014}).

\begin{figure}
\centering 
\includegraphics[scale=0.87,trim= 2.7cm 13.2cm 9cm 5.8cm]{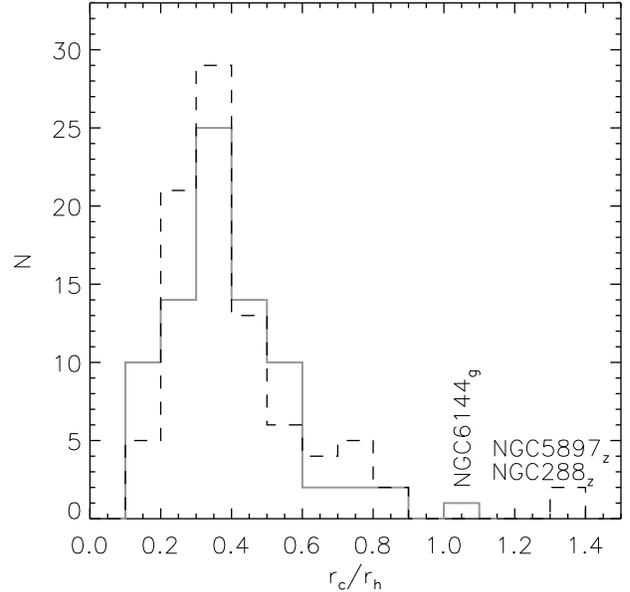}
\caption{Distribution of the ratio between the core and the half-light radii. Legend as in Fig.~\ref{fig:histrc}.} 
\label{fig:histrcoverrh}
\end{figure}

Standard dynamical models of GCs make clear predictions of the evolution of the ratio between the core and the half-light radii (see \citealt{Trenti2010,Miocchi2013} and references therein). Fig.~\ref{fig:histrcoverrh} presents the distribution of the ratio between the 
core and the half-light radii. We find a peaked distribution centred at $r_c/r_h \sim 0.4$, but do not recover the bimodal distribution found by 
\cite{Miocchi2013}, with a peak in $r_c/r_h$ at about 0.3. Both our and their $r_c/r_h$ values are in agreement with expectations from simulations of cluster dynamical evolution.

\subsection{Colour Gradients}\label{sec:colgrad}
In galaxies, colour gradients are interpreted as a metallicity gradient \citep{Tamura2000, LaBarbera2010b}.  
If colour gradients exist in GCs, these would not be linked with the metallicity, as these objects have a largely homogenous 
metallicity (although variations in light element abundances are omnipresent). However, if dynamical processes affect stellar 
populations, for instance through the creation of blue stragglers in cluster cores \citep{Ferraro2009} or stripping of red giants 
to produce AGB-manque stars \citep{Pasquato2013}, the formation of cataclysmic variables, millisecond pulsars and 
intermediate mass black holes, we might be able to detect the presence of these objects through colour variations as
a function of radius. \cite{Djorgovski1993} found that post core-collapse clusters are bluer in their centre and there were 
no colour gradients in the opposite direction. \cite{Djorgovski1991} found no blue cores in clusters with flat inner profiles, 
confirming the correlation between internal dynamics and stellar populations. This was attributed to the stripping of red giants 
in cluster cores to produce extended HBs \citep{Pasquato2013} and the formation of blue stragglers by stellar collisions.

However, \cite{Sohn1996} found colour gradients even in clusters that fitted the King profile and
suggested that internal dynamics may produce extended HBs and lead to colour gradients. 
\cite{Sohn1998} also found both red and blue cores in both normal and post core-collapse
clusters. For example, unlike \cite{Djorgovski1991}, \cite{Sohn1998} detect a red core in NGC~2808.

We find general flatness of the gradients beyond the half-light radius, which is somehow expected as we do not expect strong stellar population gradients. However, stronger gradients are apparent closer to the cluster centre. To study the colour of the cluster cores and search for 
correlations with other GC parameters, we define a new parameter:
\begin{equation}\label{eq:delta_colgrad}
\Delta_{g-z} = (g-z)_{r_c}-(g-z)_{r_h},
\end{equation}
with $(g-z)_{r_c}$ the $g-z$ colour at the average $gz$ core radius (similar for the half-light radius). This parameter is negative for blue cores and positive for red cores. 

\begin{figure}
\centering 
\includegraphics[scale=0.87,trim= 2.7cm 13.2cm 0cm 5.8cm]{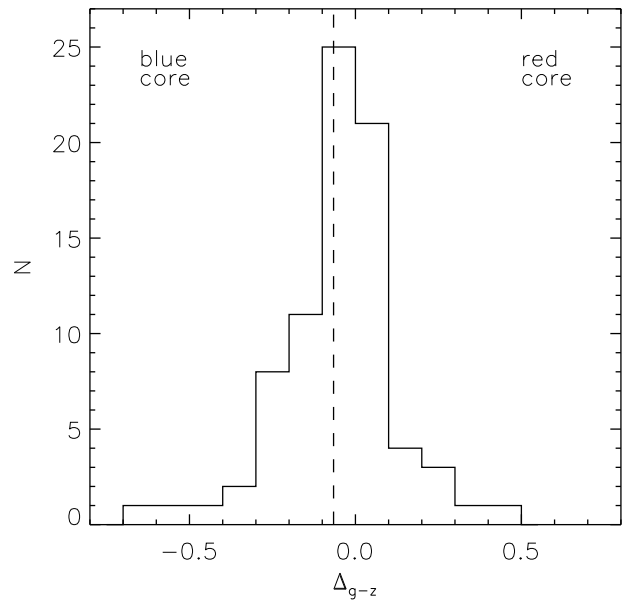}  
\caption{Distribution of $\Delta_{g-z}$ (defined in Eq.~\ref{eq:delta_colgrad}). The dashed line represents the median value of $\Delta_{g-z}$. See text for more details. }
\label{fig:histdeltagz}
\end{figure}

In Fig.~\ref{fig:histdeltagz} we present the distribution of this newly defined parameter, which is rather symmetric. 
Clusters with extremely positive values ($\Delta_{g-z}>0.3$) are NGC4833 and NGC6584, 
extremely negative values ($\Delta_{g-z}<-0.4$) are found for NGC6235, NGC6352 and NGC6681. These clusters are close to the Galactic disk or bulge, which results in SB profiles polluted by Milky Way stars and complicates the determination of the structural parameters.

Clusters in the "blue core tail" (with $-0.4 \le \Delta_{g-z} \le -0.2$) are 
NGC\,6121, NGC\,6342, NGC\,6453, NGC\,6535, NGC\,6638, NGC\,6637, NGC\,6642, NGC\,6712, NGC\,6864 and NGC\,6981. Clusters in the "red core tail" (with $0.1 \le \Delta_{g-z} \le 0.3$) are  NGC\,2298, NGC\,5634, NGC\,6171, NGC\,6273, NGC\,6341, 
NGC\,6760 and NGC\,6779. For a fraction of these clusters, we also determined the King parameters based on a SB profile centred with the RGB method, a procedure which should not be susceptible towards centrally located bright stars (see Section~\ref{sec:obsbiases}). About one third of the "tail" clusters are located towards the Galactic bulge, prohibiting an RGB-based centre determination. For a couple of clusters, only poor colour-magnitude diagrams (CMDs) were obtained and a centre determination based on the RGB was not possible.

It is generally difficult to study colour gradients, especially close to the cluster centre, because only a couple of stochastically distributed 
RGB stars can alter the SB profiles and originate colour gradients. This is especially true for low-density clusters, but can also affect clusters 
as bright as NGC~104: RGB stars and poor centring can affect the resulting King parameters drastically. 

\section{Summary}\label{sec:conclusions}

As a part of the G2C2 project, we use in the current study our optical dataset (presented in \citetalias{Vanderbeke2014a}) to derive structural parameters fitting \cite{King1962} models to the GC SB profiles. We present structural parameters in the $g$ and $z$ filters for 111 Galactic GCs, while we also include $r$ and $i$-band values for 60 clusters and $u$-band estimates for 22 GCs. For some clusters 
it was not possible to fit a representative King model to the SB profiles. Neglecting the unrealistic fits (judged by visual inspection), we present $u$, $g$, $r$, $i$ and $z$ structural parameters for 18, 94, 50, 48 and 95 GCs, respectively. Nevertheless, in order to maintain the focus on the GCs with soundly determined parameters in the discussion, we find it necessary to introduce some parameter criteria in the discussion (Section~\ref{sec:complit}). For the $u$ ($g,r,i,z$)-band, the structural parameters of 12, (80, 41, 42, 87) GCs satisfy these standards.

Because the bulk of our dataset was obtained with CTIO 0.9\,m telescope observations, suffering from a limited FOV, we extensively discuss the effects of a limited field-of-view on the derived King model parameters. In general, the resulting core radii are in good comparison with the current literature values. However, our half-light radii are slightly underestimated when compared to the literature. The concentrations (and therefore also the tidal radii) are poorly constrained, partly due to the limited radial extent of our SB profiles. 
Moreover, RGB stars are biasing the centring procedure and can provoke strong central SB cusps, which further contribute to the overestimation of the concentration parameter. The issues related to these bright stars are scrutinised based on our photometric data and simulated clusters. Logically, the effects of the randomly distributed RGB stars on the fitted King models are stronger in sparsely populated clusters, for which a star count density approach is recommended. Colour gradients are examined and can also be related to RGB stars. 

We recover the known relation between the half-light radius and the Galactocentric distance for the $g$-band, but find a lower slope for redder filters. We did not find a correlation between the scatter on this relation and other cluster properties. We find tentative evidence for a correlation between the half-light radii and [Fe/H], with metal-poor GCs being larger than metal-rich GCs. However, it turns out that this trend is due to the relation between the half-light radius and the Galactocentric distance, with metal-rich clusters being more centrally located than metal-poor clusters. 

\section*{Acknowledgments}
We thankfully acknowledge the anonymous referee for very useful comments and suggestions. JV acknowledges the support of ESO through a studentship. JV and MB acknowledge the support of the Fund for Scientific Research Flanders (FWO-Vlaanderen). The authors are grateful to CTIO for the hospitality and the dedicated assistance during the numerous observing runs. 
\bibliographystyle{aa}
\bibliography{references}

\begin{thebibliography}{92}
\expandafter\ifx\csname natexlab\endcsname\relax\def\natexlab#1{#1}\fi

\bibitem[{{Ahn} {et~al.}(2012){Ahn}, {Alexandroff}, {Allende Prieto},
  {Anderson}, {Anderton}, {Andrews}, {Aubourg}, {Bailey}, {Balbinot}, {Barnes},
  \& et~al.}]{Ahn2012}
{Ahn}, C.~P., {Alexandroff}, R., {Allende Prieto}, C., {et~al.} 2012, \apjs,
  203, 21

\bibitem[{{Alexander} \& {Gieles}(2013)}]{Alexander2013}
{Alexander}, P.~E.~R. \& {Gieles}, M. 2013, \mnras, 432, L1

\bibitem[{{Alonso-Garc{\'{\i}}a} {et~al.}(2011){Alonso-Garc{\'{\i}}a}, {Mateo},
  {Sen}, {Banerjee}, \& {von Braun}}]{AlonsoGarcia2011}
{Alonso-Garc{\'{\i}}a}, J., {Mateo}, M., {Sen}, B., {Banerjee}, M., \& {von
  Braun}, K. 2011, \aj, 141, 146

\bibitem[{{Bahcall} \& {Soneira}(1980)}]{Bahcall1980}
{Bahcall}, J.~N. \& {Soneira}, R.~M. 1980, \apjs, 44, 73

\bibitem[{{Bellazzini}(2007)}]{Bellazzini2007}
{Bellazzini}, M. 2007, \aap, 473, 171

\bibitem[{{Benacquista} \& {Downing}(2013)}]{Benacquista2013}
{Benacquista}, M.~J. \& {Downing}, J.~M.~B. 2013, Living Reviews in Relativity,
  16, 4

\bibitem[{{Bevington} \& {Robinson}(1992)}]{Bevington1992}
{Bevington}, P.~R. \& {Robinson}, D.~K. 1992, {Data reduction and error
  analysis for the physical sciences}

\bibitem[{{Bianchini} {et~al.}(2013){Bianchini}, {Varri}, {Bertin}, \&
  {Zocchi}}]{Bianchini2013}
{Bianchini}, P., {Varri}, A.~L., {Bertin}, G., \& {Zocchi}, A. 2013, \apj, 772,
  67

\bibitem[{{Brodie} \& {Strader}(2006)}]{Brodie2006}
{Brodie}, J.~P. \& {Strader}, J. 2006, \araa, 44, 193

\bibitem[{{Cohn}(1980)}]{Cohn1980}
{Cohn}, H. 1980, \apj, 242, 765

\bibitem[{{Correnti} {et~al.}(2011){Correnti}, {Bellazzini}, {Dalessandro},
  {Mucciarelli}, {Monaco}, \& {Catelan}}]{Correnti2011}
{Correnti}, M., {Bellazzini}, M., {Dalessandro}, E., {et~al.} 2011, \mnras,
  417, 2411

\bibitem[{{C{\^o}t{\'e}} {et~al.}(2001){C{\^o}t{\'e}}, {McLaughlin}, {Hanes},
  {Bridges}, {Geisler}, {Merritt}, {Hesser}, {Harris}, \& {Lee}}]{Cote2001}
{C{\^o}t{\'e}}, P., {McLaughlin}, D.~E., {Hanes}, D.~A., {et~al.} 2001, \apj,
  559, 828

\bibitem[{{de Souza} {et~al.}(2004){de Souza}, {Gadotti}, \& {dos
  Anjos}}]{Desouza2004}
{de Souza}, R.~E., {Gadotti}, D.~A., \& {dos Anjos}, S. 2004, \apjs, 153, 411

\bibitem[{{Dirsch} {et~al.}(2003){Dirsch}, {Richtler}, {Geisler}, {Forte},
  {Bassino}, \& {Gieren}}]{Dirsch2003}
{Dirsch}, B., {Richtler}, T., {Geisler}, D., {et~al.} 2003, \aj, 125, 1908

\bibitem[{{Djorgovski} \& {King}(1986)}]{Djorgovski1986}
{Djorgovski}, S. \& {King}, I.~R. 1986, \apjl, 305, L61

\bibitem[{{Djorgovski} \& {Piotto}(1993)}]{Djorgovski1993}
{Djorgovski}, S. \& {Piotto}, G. 1993, in Astronomical Society of the Pacific
  Conference Series, Vol.~50, Structure and Dynamics of Globular Clusters, ed.
  S.~G. {Djorgovski} \& G.~{Meylan}, 203

\bibitem[{{Djorgovski} {et~al.}(1991){Djorgovski}, {Piotto}, {Phinney}, \&
  {Chernoff}}]{Djorgovski1991}
{Djorgovski}, S., {Piotto}, G., {Phinney}, E.~S., \& {Chernoff}, D.~F. 1991,
  \apjl, 372, L41

\bibitem[{{Elson} {et~al.}(1987){Elson}, {Fall}, \& {Freeman}}]{Elson1987}
{Elson}, R.~A.~W., {Fall}, S.~M., \& {Freeman}, K.~C. 1987, \apj, 323, 54

\bibitem[{{Ernst} \& {Just}(2013)}]{Ernst2013}
{Ernst}, A. \& {Just}, A. 2013, \mnras, 429, 2953

\bibitem[{{Fabricius} {et~al.}(2014){Fabricius}, {Noyola}, {Rukdee}, {Saglia},
  {Bender}, {Hopp}, {Thomas}, {Opitsch}, \& {Williams}}]{Fabricius2014}
{Fabricius}, M.~H., {Noyola}, E., {Rukdee}, S., {et~al.} 2014, \apjl, 787, L26

\bibitem[{{Ferraro} \& {Lanzoni}(2009)}]{Ferraro2009}
{Ferraro}, F.~R. \& {Lanzoni}, B. 2009, in Revista Mexicana de Astronomia y
  Astrofisica Conference Series, Vol.~37, Revista Mexicana de Astronomia y
  Astrofisica Conference Series, 62--71

\bibitem[{{Ferraro} {et~al.}(1999){Ferraro}, {Paltrinieri}, {Rood}, \&
  {Dorman}}]{Ferraro1999}
{Ferraro}, F.~R., {Paltrinieri}, B., {Rood}, R.~T., \& {Dorman}, B. 1999, \apj,
  522, 983

\bibitem[{{Ferraro} {et~al.}(2003){Ferraro}, {Possenti}, {Sabbi}, {Lagani},
  {Rood}, {D'Amico}, \& {Origlia}}]{Ferraro2003}
{Ferraro}, F.~R., {Possenti}, A., {Sabbi}, E., {et~al.} 2003, \apj, 595, 179

\bibitem[{{Forbes} \& {Bridges}(2010)}]{Forbes2010}
{Forbes}, D.~A. \& {Bridges}, T. 2010, \mnras, 404, 1203

\bibitem[{{Frank} {et~al.}(2012){Frank}, {Hilker}, {Baumgardt}, {C{\^o}t{\'e}},
  {Grebel}, {Haghi}, {K{\"u}pper}, \& {Djorgovski}}]{Frank2012}
{Frank}, M.~J., {Hilker}, M., {Baumgardt}, H., {et~al.} 2012, \mnras, 423, 2917

\bibitem[{{Fregeau} {et~al.}(2002){Fregeau}, {Joshi}, {Portegies Zwart}, \&
  {Rasio}}]{Fregeau2002}
{Fregeau}, J.~M., {Joshi}, K.~J., {Portegies Zwart}, S.~F., \& {Rasio}, F.~A.
  2002, \apj, 570, 171

\bibitem[{{Fregeau} \& {Rasio}(2007)}]{Fregeau2007}
{Fregeau}, J.~M. \& {Rasio}, F.~A. 2007, \apj, 658, 1047

\bibitem[{{Fusi Pecci} {et~al.}(1992){Fusi Pecci}, {Ferraro}, {Corsi},
  {Cacciari}, \& {Buonanno}}]{FusiPecci1992}
{Fusi Pecci}, F., {Ferraro}, F.~R., {Corsi}, C.~E., {Cacciari}, C., \&
  {Buonanno}, R. 1992, \aj, 104, 1831

\bibitem[{{Gadotti}(2008)}]{Gadotti2008}
{Gadotti}, D.~A. 2008, \mnras, 384, 420

\bibitem[{{Georgiev} \& {B{\"o}ker}(2014)}]{Georgiev2014}
{Georgiev}, I.~Y. \& {B{\"o}ker}, T. 2014, \mnras, 441, 3570

\bibitem[{{Gieles} {et~al.}(2010){Gieles}, {Baumgardt}, {Heggie}, \&
  {Lamers}}]{Gieles2010}
{Gieles}, M., {Baumgardt}, H., {Heggie}, D.~C., \& {Lamers}, H.~J.~G.~L.~M.
  2010, \mnras, 408, L16

\bibitem[{{Goldsbury} {et~al.}(2013){Goldsbury}, {Heyl}, \&
  {Richer}}]{Goldsbury2013}
{Goldsbury}, R., {Heyl}, J., \& {Richer}, H. 2013, \apj, 778, 57

\bibitem[{{Grillmair} {et~al.}(1995){Grillmair}, {Freeman}, {Irwin}, \&
  {Quinn}}]{Grillmair1995}
{Grillmair}, C.~J., {Freeman}, K.~C., {Irwin}, M., \& {Quinn}, P.~J. 1995, \aj,
  109, 2553

\bibitem[{{Harris}(1996)}]{Harris1996}
{Harris}, W.~E. 1996, \aj, 112, 1487

\bibitem[{{Harris}(2009)}]{Harris2009}
{Harris}, W.~E. 2009, \apj, 699, 254

\bibitem[{{Heggie} \& {Hut}(2003)}]{Heggie2003}
{Heggie}, D. \& {Hut}, P. 2003, {The Gravitational Million-Body Problem: A
  Multidisciplinary Approach to Star Cluster Dynamics}

\bibitem[{{Hernandez} {et~al.}(2013){Hernandez}, {Jim{\'e}nez}, \&
  {Allen}}]{Hernandez2013}
{Hernandez}, X., {Jim{\'e}nez}, M.~A., \& {Allen}, C. 2013, \mnras, 428, 3196

\bibitem[{{Hurley} \& {Shara}(2012)}]{Hurley2012}
{Hurley}, J.~R. \& {Shara}, M.~M. 2012, \mnras, 425, 2872

\bibitem[{{Jord{\'a}n}(2004)}]{Jordan2004}
{Jord{\'a}n}, A. 2004, \apjl, 613, L117

\bibitem[{{Jord{\'a}n} {et~al.}(2005){Jord{\'a}n}, {C{\^o}t{\'e}}, {Blakeslee},
  {Ferrarese}, {McLaughlin}, {Mei}, {Peng}, {Tonry}, {Merritt},
  {Milosavljevi{\'c}}, {Sarazin}, {Sivakoff}, \& {West}}]{Jordan2005}
{Jord{\'a}n}, A., {C{\^o}t{\'e}}, P., {Blakeslee}, J.~P., {et~al.} 2005, \apj,
  634, 1002

\bibitem[{{Jord{\'a}n} {et~al.}(2009){Jord{\'a}n}, {Peng}, {Blakeslee},
  {C{\^o}t{\'e}}, {Eyheramendy}, {Ferrarese}, {Mei}, {Tonry}, \&
  {West}}]{Jordan2009}
{Jord{\'a}n}, A., {Peng}, E.~W., {Blakeslee}, J.~P., {et~al.} 2009, \apjs, 180,
  54

\bibitem[{{Jordi} \& {Grebel}(2010)}]{Jordi2010}
{Jordi}, K. \& {Grebel}, E.~K. 2010, \aap, 522, A71

\bibitem[{{Kacharov} {et~al.}(2014){Kacharov}, {Bianchini}, {Koch}, {Frank},
  {Martin}, {van de Ven}, {Puzia}, {McDonald}, {Johnson}, \&
  {Zijlstra}}]{Kacharov2014}
{Kacharov}, N., {Bianchini}, P., {Koch}, A., {et~al.} 2014, \aap, 567, A69

\bibitem[{{King}(1962)}]{King1962}
{King}, I. 1962, \aj, 67, 471

\bibitem[{{King}(1966)}]{King1966}
{King}, I.~R. 1966, \aj, 71, 64

\bibitem[{{Kissler-Patig} {et~al.}(1997){Kissler-Patig}, {Kohle}, {Hilker},
  {Richtler}, {Infante}, \& {Quintana}}]{KisslerPatig1997}
{Kissler-Patig}, M., {Kohle}, S., {Hilker}, M., {et~al.} 1997, \aap, 319, 470

\bibitem[{{Kundu} \& {Whitmore}(1998)}]{Kundu1998}
{Kundu}, A. \& {Whitmore}, B.~C. 1998, \aj, 116, 2841

\bibitem[{{Kundu} {et~al.}(1999){Kundu}, {Whitmore}, {Sparks}, {Macchetto},
  {Zepf}, \& {Ashman}}]{Kundu1999}
{Kundu}, A., {Whitmore}, B.~C., {Sparks}, W.~B., {et~al.} 1999, \apj, 513, 733

\bibitem[{{La Barbera} {et~al.}(2010){La Barbera}, {De Carvalho}, {De La Rosa},
  {Gal}, {Swindle}, \& {Lopes}}]{LaBarbera2010b}
{La Barbera}, F., {De Carvalho}, R.~R., {De La Rosa}, I.~G., {et~al.} 2010,
  \aj, 140, 1528

\bibitem[{{Larsen} \& {Brodie}(2003)}]{Larsen2003}
{Larsen}, S.~S. \& {Brodie}, J.~P. 2003, \apj, 593, 340

\bibitem[{{Larsen} {et~al.}(2001){Larsen}, {Brodie}, {Huchra}, {Forbes}, \&
  {Grillmair}}]{Larsen2001}
{Larsen}, S.~S., {Brodie}, J.~P., {Huchra}, J.~P., {Forbes}, D.~A., \&
  {Grillmair}, C.~J. 2001, \aj, 121, 2974

\bibitem[{{Lee} {et~al.}(1998){Lee}, {Kim}, \& {Geisler}}]{Lee1998}
{Lee}, M.~G., {Kim}, E., \& {Geisler}, D. 1998, \aj, 115, 947

\bibitem[{{Lin} {et~al.}(2013){Lin}, {Irwin}, {Webb}, {Barret}, \&
  {Remillard}}]{Lin2013}
{Lin}, D., {Irwin}, J.~A., {Webb}, N.~A., {Barret}, D., \& {Remillard}, R.~A.
  2013, \apj, 779, 149

\bibitem[{{L{\"u}tzgendorf} {et~al.}(2013){L{\"u}tzgendorf}, {Kissler-Patig},
  {Gebhardt}, {Baumgardt}, {Noyola}, {de Zeeuw}, {Neumayer}, {Jalali}, \&
  {Feldmeier}}]{Lutzgendorf2013}
{L{\"u}tzgendorf}, N., {Kissler-Patig}, M., {Gebhardt}, K., {et~al.} 2013,
  \aap, 552, A49

\bibitem[{{Mackey} \& {Gilmore}(2003)}]{Mackey2003}
{Mackey}, A.~D. \& {Gilmore}, G.~F. 2003, \mnras, 338, 85

\bibitem[{{Mackey} \& {van den Bergh}(2005)}]{Mackey2005}
{Mackey}, A.~D. \& {van den Bergh}, S. 2005, \mnras, 360, 631

\bibitem[{{Mackey} {et~al.}(2008){Mackey}, {Wilkinson}, {Davies}, \&
  {Gilmore}}]{Mackey2008}
{Mackey}, A.~D., {Wilkinson}, M.~I., {Davies}, M.~B., \& {Gilmore}, G.~F. 2008,
  \mnras, 386, 65

\bibitem[{{Masters} {et~al.}(2010){Masters}, {Jord{\'a}n}, {C{\^o}t{\'e}},
  {Ferrarese}, {Blakeslee}, {Infante}, {Peng}, {Mei}, \& {West}}]{Masters2010}
{Masters}, K.~L., {Jord{\'a}n}, A., {C{\^o}t{\'e}}, P., {et~al.} 2010, \apj,
  715, 1419

\bibitem[{{McLaughlin}(2000)}]{McLaughlin2000}
{McLaughlin}, D.~E. 2000, \apj, 539, 618

\bibitem[{{McLaughlin} \& {van der Marel}(2005)}]{McLaughlin2005}
{McLaughlin}, D.~E. \& {van der Marel}, R.~P. 2005, \apjs, 161, 304

\bibitem[{{Miocchi} {et~al.}(2013){Miocchi}, {Lanzoni}, {Ferraro},
  {Dalessandro}, {Vesperini}, {Pasquato}, {Beccari}, {Pallanca}, \&
  {Sanna}}]{Miocchi2013}
{Miocchi}, P., {Lanzoni}, B., {Ferraro}, F.~R., {et~al.} 2013, \apj, 774, 151

\bibitem[{{Newell} \& {Oneil}(1978)}]{Newell1978}
{Newell}, B. \& {Oneil}, Jr., E.~J. 1978, \apjs, 37, 27

\bibitem[{{Noyola} \& {Gebhardt}(2006)}]{Noyola2006}
{Noyola}, E. \& {Gebhardt}, K. 2006, \aj, 132, 447

\bibitem[{{Odenkirchen} {et~al.}(2001){Odenkirchen}, {Grebel}, {Rockosi},
  {Dehnen}, {Ibata}, {Rix}, {Stolte}, {Wolf}, {Anderson}, {Bahcall},
  {Brinkmann}, {Csabai}, {Hennessy}, {Hindsley}, {Ivezi{\'c}}, {Lupton},
  {Munn}, {Pier}, {Stoughton}, \& {York}}]{Odenkirchen2001}
{Odenkirchen}, M., {Grebel}, E.~K., {Rockosi}, C.~M., {et~al.} 2001, \apjl,
  548, L165

\bibitem[{{Olszewski} {et~al.}(2009){Olszewski}, {Saha}, {Knezek},
  {Subramaniam}, {de Boer}, \& {Seitzer}}]{Olszewski2009}
{Olszewski}, E.~W., {Saha}, A., {Knezek}, P., {et~al.} 2009, \aj, 138, 1570

\bibitem[{{Pasquato} {et~al.}(2013){Pasquato}, {Raimondo}, {Brocato}, {Chung},
  {Moraghan}, \& {Lee}}]{Pasquato2013}
{Pasquato}, M., {Raimondo}, G., {Brocato}, E., {et~al.} 2013, \aap, 554, A129

\bibitem[{{Peng} {et~al.}(2002){Peng}, {Ho}, {Impey}, \& {Rix}}]{Peng2002}
{Peng}, C.~Y., {Ho}, L.~C., {Impey}, C.~D., \& {Rix}, H.-W. 2002, \aj, 124, 266

\bibitem[{{Peng} {et~al.}(2010){Peng}, {Ho}, {Impey}, \& {Rix}}]{Peng2010}
{Peng}, C.~Y., {Ho}, L.~C., {Impey}, C.~D., \& {Rix}, H.-W. 2010, \aj, 139,
  2097

\bibitem[{{Press} \& {Schechter}(1974)}]{Press1974}
{Press}, W.~H. \& {Schechter}, P. 1974, \apj, 187, 425

\bibitem[{{Puzia} {et~al.}(1999){Puzia}, {Kissler-Patig}, {Brodie}, \&
  {Huchra}}]{Puzia1999}
{Puzia}, T.~H., {Kissler-Patig}, M., {Brodie}, J.~P., \& {Huchra}, J.~P. 1999,
  \aj, 118, 2734

\bibitem[{{Puzia} {et~al.}(2014){Puzia}, {Paolillo}, {Goudfrooij}, {Maccarone},
  {Fabbiano}, \& {Angelini}}]{Puzia2014}
{Puzia}, T.~H., {Paolillo}, M., {Goudfrooij}, P., {et~al.} 2014, \apj, 786, 78

\bibitem[{{Salinas} {et~al.}(2012){Salinas}, {J{\'{\i}}lkov{\'a}}, {Carraro},
  {Catelan}, \& {Amigo}}]{Salinas2012}
{Salinas}, R., {J{\'{\i}}lkov{\'a}}, L., {Carraro}, G., {Catelan}, M., \&
  {Amigo}, P. 2012, \mnras, 421, 960

\bibitem[{{Schlafly} \& {Finkbeiner}(2011)}]{Schlafly2011}
{Schlafly}, E.~F. \& {Finkbeiner}, D.~P. 2011, \apj, 737, 103

\bibitem[{{Simunovic} \& {Puzia}(2014)}]{Simunovic2013}
{Simunovic}, M. \& {Puzia}, T.~H. 2014, \apj, 782, 49

\bibitem[{{Smith} {et~al.}(2002){Smith}, {Tucker}, {Kent}, {Richmond},
  {Fukugita}, {Ichikawa}, {Ichikawa}, {Jorgensen}, {Uomoto}, {Gunn}, {Hamabe},
  {Watanabe}, {Tolea}, {Henden}, {Annis}, {Pier}, {McKay}, {Brinkmann}, {Chen},
  {Holtzman}, {Shimasaku}, \& {York}}]{Smith2002}
{Smith}, J.~A., {Tucker}, D.~L., {Kent}, S., {et~al.} 2002, \aj, 123, 2121

\bibitem[{{Sohn} {et~al.}(1996){Sohn}, {Byun}, \& {Chun}}]{Sohn1996}
{Sohn}, Y.-J., {Byun}, Y.-I., \& {Chun}, M.-S. 1996, \apss, 243, 379

\bibitem[{{Sohn} {et~al.}(1998){Sohn}, {Byun}, {Yim}, {Rhee}, \&
  {Chun}}]{Sohn1998}
{Sohn}, Y.-J., {Byun}, Y.-I., {Yim}, H.-S., {Rhee}, M.-H., \& {Chun}, M.-S.
  1998, Journal of Astronomy and Space Sciences, 15, 1

\bibitem[{{Sollima} {et~al.}(2011){Sollima}, {Mart{\'{\i}}nez-Delgado},
  {Valls-Gabaud}, \& {Pe{\~n}arrubia}}]{Sollima2011}
{Sollima}, A., {Mart{\'{\i}}nez-Delgado}, D., {Valls-Gabaud}, D., \&
  {Pe{\~n}arrubia}, J. 2011, \apj, 726, 47

\bibitem[{{Stetson}(1987)}]{Stetson1987}
{Stetson}, P.~B. 1987, \pasp, 99, 191

\bibitem[{{Tamura} \& {Ohta}(2000)}]{Tamura2000}
{Tamura}, N. \& {Ohta}, K. 2000, \aj, 120, 533

\bibitem[{{Trager} {et~al.}(1995){Trager}, {King}, \&
  {Djorgovski}}]{Trager1995}
{Trager}, S.~C., {King}, I.~R., \& {Djorgovski}, S. 1995, \aj, 109, 218

\bibitem[{{Trenti} {et~al.}(2010){Trenti}, {Vesperini}, \&
  {Pasquato}}]{Trenti2010}
{Trenti}, M., {Vesperini}, E., \& {Pasquato}, M. 2010, \apj, 708, 1598

\bibitem[{{van den Bergh}(2012)}]{Vandenbergh2012}
{van den Bergh}, S. 2012, \apj, 746, 189

\bibitem[{{van den Bergh} {et~al.}(1991){van den Bergh}, {Morbey}, \&
  {Pazder}}]{Vandenbergh1991}
{van den Bergh}, S., {Morbey}, C., \& {Pazder}, J. 1991, \apj, 375, 594

\bibitem[{{VandenBerg} {et~al.}(2013){VandenBerg}, {Brogaard}, {Leaman}, \&
  {Casagrande}}]{VandenBerg2013}
{VandenBerg}, D.~A., {Brogaard}, K., {Leaman}, R., \& {Casagrande}, L. 2013,
  \apj, 775, 134

\bibitem[{{Vanderbeke} {et~al.}(2014{\natexlab{a}}){Vanderbeke}, {West}, {De
  Propris}, {Peng}, {Blakeslee}, {Jord{\'a}n}, {C{\^o}t{\'e}}, {Gregg},
  {Ferrarese}, {Takamiya}, \& {Baes}}]{Vanderbeke2014a}
{Vanderbeke}, J., {West}, M.~J., {De Propris}, R., {et~al.} 2014{\natexlab{a}},
  \mnras, 437, 1725

\bibitem[{{Vanderbeke} {et~al.}(2014{\natexlab{b}}){Vanderbeke}, {West}, {De
  Propris}, {Peng}, {Blakeslee}, {Jord{\'a}n}, {C{\^o}t{\'e}}, {Gregg},
  {Ferrarese}, {Takamiya}, \& {Baes}}]{Vanderbeke2014b}
{Vanderbeke}, J., {West}, M.~J., {De Propris}, R., {et~al.} 2014{\natexlab{b}},
  \mnras, 437, 1734

\bibitem[{{Vesperini} \& {Chernoff}(1994)}]{Vesperini1994}
{Vesperini}, E. \& {Chernoff}, D.~F. 1994, \apj, 431, 231

\bibitem[{{Vesperini} \& {Trenti}(2010)}]{Vesperini2010}
{Vesperini}, E. \& {Trenti}, M. 2010, \apjl, 720, L179

\bibitem[{{Wang} \& {Ma}(2013)}]{Wang2013}
{Wang}, S. \& {Ma}, J. 2013, \aj, 146, 20

\bibitem[{{Wilson}(1975)}]{Wilson1975}
{Wilson}, C.~P. 1975, \aj, 80, 175

\bibitem[{{York} {et~al.}(2000){York}, {Adelman}, {Anderson}, {Anderson},
  {Annis}, {Bahcall}, {Bakken}, {Barkhouser}, {Bastian}, {Berman}, {Boroski},
  {Bracker}, {Briegel}, {Briggs}, {Brinkmann}, {Brunner}, {Burles}, {Carey},
  {Carr}, {Castander}, {Chen}, {Colestock}, {Connolly}, {Crocker}, {Csabai},
  {Czarapata}, {Davis}, {Doi}, {Dombeck}, {Eisenstein}, {Ellman}, {Elms},
  {Evans}, {Fan}, {Federwitz}, {Fiscelli}, {Friedman}, {Frieman}, {Fukugita},
  {Gillespie}, {Gunn}, {Gurbani}, {de Haas}, {Haldeman}, {Harris}, {Hayes},
  {Heckman}, {Hennessy}, {Hindsley}, {Holm}, {Holmgren}, {Huang}, {Hull},
  {Husby}, {Ichikawa}, {Ichikawa}, {Ivezi{\'c}}, {Kent}, {Kim}, {Kinney},
  {Klaene}, {Kleinman}, {Kleinman}, {Knapp}, {Korienek}, {Kron}, {Kunszt},
  {Lamb}, {Lee}, {Leger}, {Limmongkol}, {Lindenmeyer}, {Long}, {Loomis},
  {Loveday}, {Lucinio}, {Lupton}, {MacKinnon}, {Mannery}, {Mantsch}, {Margon},
  {McGehee}, {McKay}, {Meiksin}, {Merelli}, {Monet}, {Munn}, {Narayanan},
  {Nash}, {Neilsen}, {Neswold}, {Newberg}, {Nichol}, {Nicinski}, {Nonino},
  {Okada}, {Okamura}, {Ostriker}, {Owen}, {Pauls}, {Peoples}, {Peterson},
  {Petravick}, {Pier}, {Pope}, {Pordes}, {Prosapio}, {Rechenmacher}, {Quinn},
  {Richards}, {Richmond}, {Rivetta}, {Rockosi}, {Ruthmansdorfer}, {Sandford},
  {Schlegel}, {Schneider}, {Sekiguchi}, {Sergey}, {Shimasaku}, {Siegmund},
  {Smee}, {Smith}, {Snedden}, {Stone}, {Stoughton}, {Strauss}, {Stubbs},
  {SubbaRao}, {Szalay}, {Szapudi}, {Szokoly}, {Thakar}, {Tremonti}, {Tucker},
  {Uomoto}, {Vanden Berk}, {Vogeley}, {Waddell}, {Wang}, {Watanabe},
  {Weinberg}, {Yanny}, {Yasuda}, \& {SDSS Collaboration}}]{York2000}
{York}, D.~G., {Adelman}, J., {Anderson}, Jr., J.~E., {et~al.} 2000, \aj, 120,
  1579

\end{thebibliography}

\end{document}